\documentclass[twocolumn,english,superscriptaddress,prb]{revtex4-2}
\usepackage[T1]{fontenc}
\usepackage[latin9]{inputenc}
\usepackage{graphicx}
\usepackage{amsmath}
\usepackage[colorlinks = true,
	linkcolor = blue,
	urlcolor  = blue,
	citecolor = blue,
	anchorcolor = blue]{hyperref}
\usepackage[usenames, dvipsnames]{color}

\makeatletter
\usepackage{babel}
\usepackage{float}

\usepackage{tabularx}
\newcolumntype{C}{>{\centering\arraybackslash}X}

\makeatother

\begin{document}


%
%

\title{Beyond magnons in  $\rm Nd_2ScNbO_7$: An Ising pyrochlore antiferromagnet with all in all out order and random fields}

\author{A. Scheie}
\address{Neutron Scattering Division, Oak Ridge National Laboratory, Oak Ridge, Tennessee 37831, USA}
\address{Institute for Quantum Matter and Department of Physics and Astronomy, Johns Hopkins University, Baltimore, MD 21218}


\author{M. Sanders}
\address{Department of Chemistry, Princeton University, Princeton, NJ 08544 }

\author{Yiming Qiu}
\address{NIST Center for Neutron Research, National Institute of Standards and Technology, Gaithersburg, MD 20899}

\author{T.R. Prisk}
\address{NIST Center for Neutron Research, National Institute of Standards and Technology, Gaithersburg, MD 20899}

\author{R.J. Cava}
\address{Department of Chemistry, Princeton University, Princeton, NJ 08544 }

\author{C. Broholm}
\address{Institute for Quantum Matter and Department of Physics and Astronomy, Johns Hopkins University, Baltimore, MD 21218}
\address{NIST Center for Neutron Research, National Institute of Standards and Technology, Gaithersburg, MD 20899}
\address{Department of Materials Science and Engineering, Johns Hopkins University, Baltimore, MD 21218}

\date{\today}

\begin{abstract}
We report the low temperature magnetic properties of Nd$^{3+}$ pyrochlore $\rm Nd_2ScNbO_7$. Susceptibility and magnetization show an easy-axis moment, and heat capacity reveals a phase transition to long range order at $T_N=371(2)$~mK with a fully recovered $\Delta S = R \ln(2)$, 53\% of it recovered for $T>T_N$. Elastic neutron scattering shows a long-range all-in-all-out magnetic order with low-$Q$ diffuse elastic scattering. Inelastic neutron scattering shows a low-energy flat-band, indicating a magnetic Hamiltonian similar to $\rm Nd_2Zr_2O_7$. Nuclear hyperfine excitations measured by ultra-high-resolution neutron backscattering indicates a distribution of static electronic moments below $T_N$, which may be due to B-site disorder influencing Nd crystal electric fields.
Analysis of heat capacity data shows an unexpected $T$-linear or $T^{3/2}$ term which is inconsistent with conventional magnon quasiparticles, but is consistent with fractionalized spinons or gapless local spin excitations. We use legacy data to show similar behavior in $\rm Nd_2Zr_2O_7$. Comparing local static moments also reveals a suppression of the nuclear Schottky anomaly in temperature, evidencing a fraction of Nd sites with nearly zero static moment, consistent with exchange-disorder-induced random singlet formation. 
Taken together, these measurements suggest an unusual fluctuating magnetic ground state which mimics a spin-liquid---but may not actually be one.
\end{abstract}

\maketitle

\section{Introduction}

Quantum spin liquids are a long-sought but elusive state of matter in which magnetic spins form a many-body entangled state \cite{Balents2010review,Savary_2016review,broholm2019quantum,Knolle2019_review}. The quantum fluctuating state supports emergent topological quasiparticles with unusual properties, like magnetic charge, fractionalized values of electron spin, and non-locality \cite{SpinIce_review,broholm2019quantum,Knolle2019_review,Rau_2019_review}. Although the quest for a definitive 3D quantum spin liquid has been unsuccessful so far \cite{Knolle2019_review,broholm2019quantum}, there are proposals that the features of interest---exotic quasiparticles---appear even in materials which have long-range magnetic order but magnetic exchange Hamiltonians close to a quantum spin liquid (called "proximate" or "condensed" quantum spin liquids) \cite{Rousochatzakis_2019,banerjee2016proximate,chern2018magnetic,Liu_2019}. Such behavior has been observed in 1D spin chains \cite{Tennant_1993,Tennant_1995}, and is theorized to exist for higher dimensional materials as well. Thus, even magnetically ordered materials can be relevant to the search for exotic quasiparticles of quantum spin liquids.

The pyrochlore lattice, with its magnetically frustrating geometry of corner-sharing tetrahedra, features prominently in proposals for spin-liquids \cite{Rau_2019_review,SpinIce_review,Yan2017,Canals1998,Savary_2012,Hermele_2004,Brooks-Bartlett_2014} and proximate spin liquids \cite{Liu_2019,chern2018magnetic}.
A recent class of compounds that has received much attention is pyrochlore magnets based on the  Nd$^{3+}$ ion \cite{Anand_2015,Anand_2017,Benton_2016,Bertin_2015,Hatnean_2015,Lhotel_2015,Xu_2016_musr,Petit2016,Opherden_2017,Lhotel2018,Xu_2018_Field,Xu_2019_Anisotropic,xu2019order,mauws2019order,Xu_2015,leger2021spin}. $\rm Nd_2Zr_2O_7$, $\rm Nd_2Hf_2O_7$, and $\rm Nd_2ScNbO_7$ show many similar features which defy conventional expectations: a pinch-point in the $Q$-dependence of the inelastic magnetic neutron scattering cross section  \cite{Petit2016,Xu_2019_Anisotropic} and a strongly reduced magnetic ordered moment \cite{Anand_2015,Lhotel_2015}.
The behavior of these materials has been attributed to "moment fragmentation" \cite{Petit2016}: a theoretically proposed magnetic state with a crystallized lattice of magnetic monopoles, forming a three-in-one-out order on a pyrochlore lattice \cite{Brooks-Bartlett_2014,Hermele_2014,Lefrancois2017fragmented,Cathelin2020fragmented}. 
However, despite much use of the fragmentation language \cite{Benton_2016,mauws2019order}, the three-in-one-out order is absent from these materials and thus it is not the ground state but the excitations which are considered fragmented \cite{Lhotel2020review}.

In this article we focus on $\rm Nd_2ScNbO_7$, a recently reported Nd$^{3+}$ pryochlore with a disordered Nb and Sc $B$-site lattice \cite{sanders2017thesis,mauws2019order}. We use magnetization, susceptibility, heat capacity, and elastic, quasielastic, and inelastic neutron scattering to characterize its ground state magnetism and show that $\rm Nd_2ScNbO_7$ exhibits many of the same behaviors as the sister-compounds $\rm Nd_2Zr_2O_7$ and $\rm Nd_2Hf_2O_7$. We report an all-in-all-out magnetic order with easy-axis anisotropy and a distribution of static ordered moments possibly due to disordered crystal-electric fields. Comparing measures of the local ordered moment magnitude indicates some sites with no static moment, and analysis of low-temperature heat capacity reveals an unexpected density of states. 
We offer two possible explanations of these results: a moment modulated quantum fragmented phase with exotic quasiparticles, or local disorder-induced spin singlets within the long-range ordered phase, possibly a result of proximity to a spin-liquid phase.
We then use these results and analysis of previously published  $\rm Nd_2Zr_2O_7$ data to show that this same behavior appears in other Nd$^{3+}$ pyrochlores, indicating a surprising similarity in multiple members of the Nd pyrochlore family.

\section{Experiments and Results}

\subsection{Sample Synthesis}

Polycrystaline $\rm Nd_2ScNbO_7$ was synthesized from a stoichiometric mixture of $\rm Nd_2O_3$, $\rm Sc_2O_3$, and $\rm Nb_2O_5$, which we mixed and pre-reacted at 1000 \textdegree C for 60 hours with repeated grinding and pelletizing.
We confirmed the pyrochlore crystal structure with a Bruker D8 Advance Eco x-ray diffractometer with Cu K$\alpha$ radiation ($\lambda = 1.5418$ \AA) and a Lynxeye detector. Refinements showed the lattice parameter $a=b=c=10.533(6)$~\AA~ at 296~K, and equal mixing of Nb and Sc on the B site of the pyrochlore lattice. Further details are in ref. \cite{sanders2017thesis}.

\subsection{Susceptibility and Magnetization}

\begin{figure}
	\centering\includegraphics[scale=0.6]{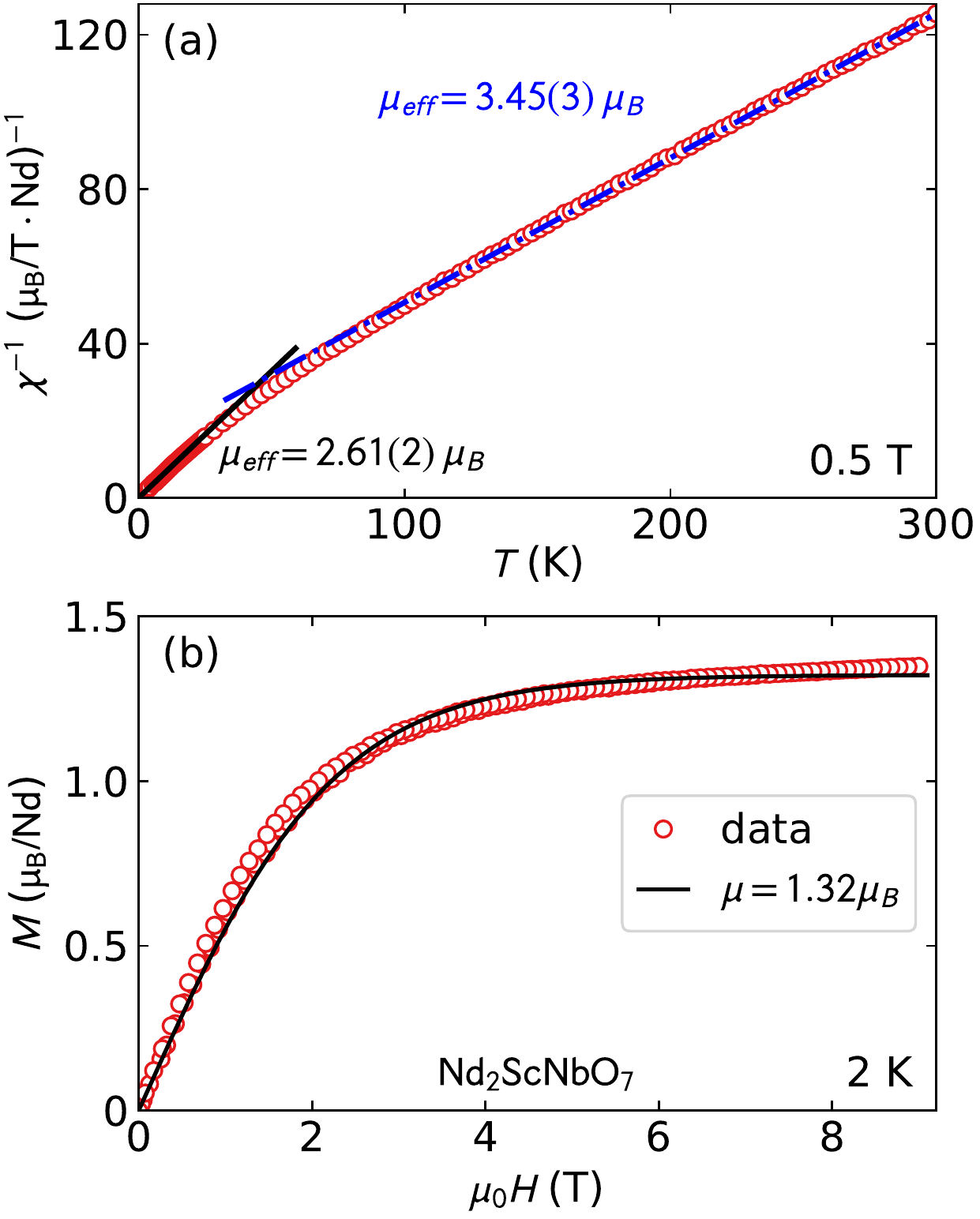}
	
	\caption{(a) Inverse susceptibility of powder $\rm{Nd_2ScNbO_7}$ and CW fits, showing high temperature and low temperature regimes where the ground state crystal field doublet is thermally populated.  (b) Powder averaged magnetization at 2 K, showing a saturation magnetization of roughly 1.3 $\mu_B$.}

	\label{flo:suscep_mag}
	
\end{figure}

We measured the susceptibility and magnetization of a pressed pellet sample using a Quantum Design Physical Properties Measurement System (PPMS) \cite{NIST_disclaimer}. Magnetic susceptibility was measured between $T=1.8$~K and $T=300$~K at 0.5 T, and magnetization at $T=2$~K was measured between 0~T and 9~T. The data and fits are shown in Fig. \ref{flo:suscep_mag}.
The temperature dependence of the susceptibility shows a bend at $T=50$~K typical of rare-earth ions with low-lying CEF levels. 
We fit the susceptibility data to a Curie-Weiss law at high temperatures (100~K$<T<$300~K) and low temperatures (2~K$<T<$10~K). The high-temperature Curie-Weiss fit yields $\mu_{eff}=3.45(3)\>{\rm \mu_B}$, close to the expected free ion value of $g_J {\rm \mu_B} \sqrt{J(J+1)} = 3.618~{\rm \mu_B}$ where $J=9/2$ and $g_j=8/11$. . The low-temperature fit (sensitive to the lowest CEF doublet and thus the ground state magnetism) yields $\mu_{eff}=2.61(2)~{\rm \mu_B}$ and $\Theta_{cw} = -0.07(4)$~K, so the net magnetic interactions nearly cancel out but are slightly antiferromagnetic. 

We fit the magnetization versus applied field for the pressed pellet to a Brillouin function (Fig. \ref{flo:suscep_mag}(b)) which indicates a saturated moment of 1.323(5) $\mu_B$. This value is almost exactly half of the free-ion saturation moment, which is consistent with powder averaging of  an easy-axis magnet (typical for Nd on the pyrochlore lattice \cite{Scheie2018_CEF,Anand_2017,Lhotel_2015}). The  slight deviations from the Brillouin function lineshape are presumably due to magnetic exchange and higher crystal field levels. 

\subsection{Heat Capacity}

\begin{figure}
	\centering\includegraphics[scale=0.6]{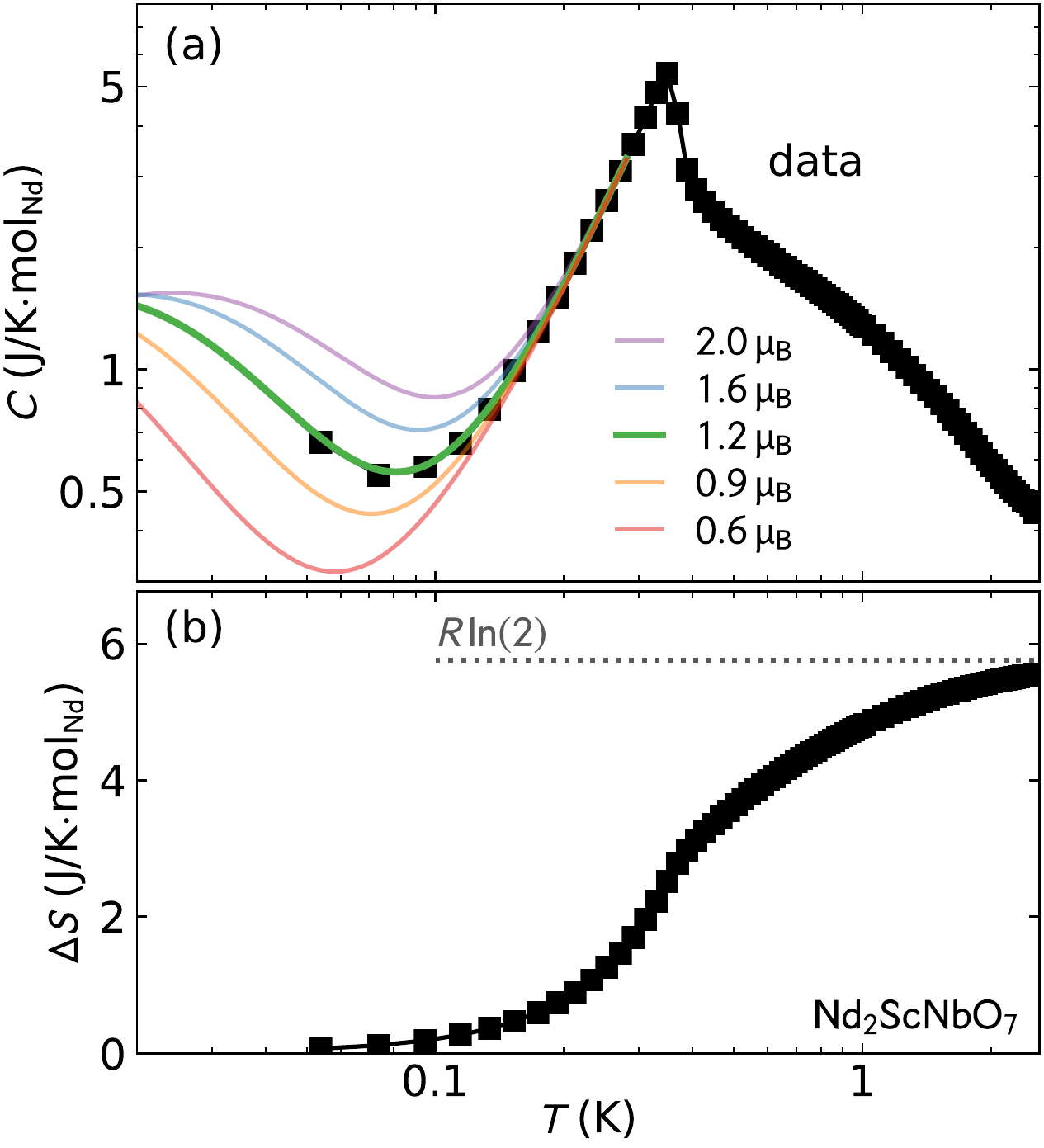}
	
	\caption{(a) Low temperature heat capacity of $\rm{Nd_2ScNbO_7}$, showing a phase transition at 370 mK. The colored lines show calculated nuclear Schottky anomalies for different ordered electronic Nd moment sizes. (b) Entropy calculated from heat capacity after subtracting the nuclear Schottky term.}
	\label{flo:heatcapacity}
	
\end{figure}

We measured zero-field heat capacity of $\rm Nd_2ScNbO_7$ as a function of temperature using a Quantum Design PPMS with a dilution refrigerator insert \cite{NIST_disclaimer}. Heat capacity at each temperature was measured three times with the semi-adiabatic relaxation method, and then averaged. The sample was a cold-pressed 1.3 mg powder pellet with equal masses of $\rm Nd_2ScNbO_7$ and silver powder for thermal connection; we subtracted the heat capacity of silver (measured separately) from the raw data.  The data are shown in Fig. \ref{flo:heatcapacity}, and reveal a peak at 370(10) mK indicative of a magnetic phase transition.

The heat capacity below 100 mK shows an upturn that we associate with a nuclear Schottky anomaly (where the Nd nuclear moments become polarized along the local field-direction of the static electronic magnetism). We fit this Schottky anomaly using PyNuclearSchottky \cite{PyNuclearSchottky} (which uses the hyperfine coupling constants in ref. \cite{Bleaney1963}) to simulate the nuclear heat capacity for Nd$^{3+}$ and a phenomenological $C_{electronic} = a T^n + \gamma T$ for the low-temperature tail of the lambda anomaly. These fits yield a sample averaged  ordered moment of $\langle \mu \rangle = 1.171(13) {\rm \mu_B}$ and electronic exponent $n=2.20(2)$. (Possible reasons for the non-integer exponent are discussed in section \ref{sec:Analysis}.) As Fig. \ref{flo:heatcapacity}(a) shows, the ordered moment is well-constrained by the high-temperature Schottky tail, even though the entire peak was not measured.

We calculated the entropy recovered across the phase transition $\Delta S = \int \frac{C}{T}dT$ after subtracting the nuclear Schottky anomaly. As Fig. \ref{flo:heatcapacity}(b) shows, the electronic magnetic entropy across the entire temperature range probed converges to $R \ln2$, the entropy of a Kramers doublet.

\subsection{Neutron Scattering}

We performed two neutron experiments on $\rm Nd_2ScNbO_7$: one experiment using the MACS triple axis spectrometer at the NCNR to measure the diffraction pattern and the spin excitations, and one experiment using the HFBS backscattering spectrometer at the NCNR to measure the nuclear hyperfine excitations.
Both experiments were performed on the same 9.33 g powder sample sealed under 10 bar He in a copper can, and mounted in a dilution refrigerator.

\subsubsection{MACS experiment}

\begin{figure}
	\centering\includegraphics[width=0.47\textwidth]{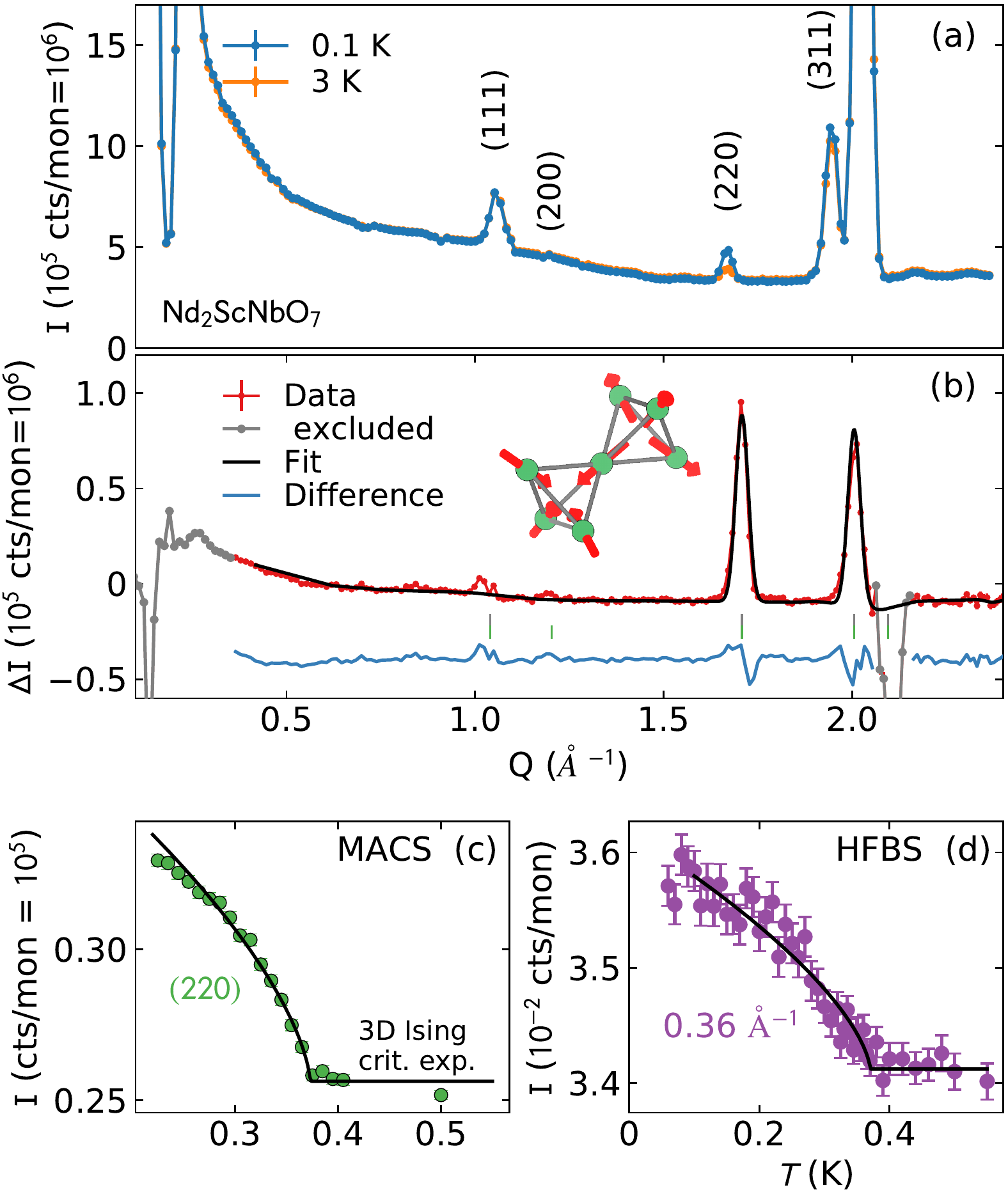}
	
	\caption{(a) Elastic neutron scattering of $\rm Nd_2ScNbO_7$ at 0.1 K and 3 K, showing significant temperature dependence on the (220) and (311) peaks. (b) Rietveld refinement of temperature subtracted scattering data, showing a best fit all-in-all-out structure. Note also the temperature dependent low-Q diffuse scattering. (c) Energy-integrated $\rm Nd_2ScNbO_7$ scattering as a function of temperature at the (220) Bragg peak. (d) Elastic ($\pm 0.4 \> {\rm \mu eV}$) low-$Q$ scattering as a function of temperature. Both (c) and (d) order parameter curves are consistent with a 3D Ising critical exponent, shown in black. Error bars indicate one standard deviation.}
	\label{flo:refinement}
	
\end{figure}

\begin{figure}
	\centering\includegraphics[width=0.47\textwidth]{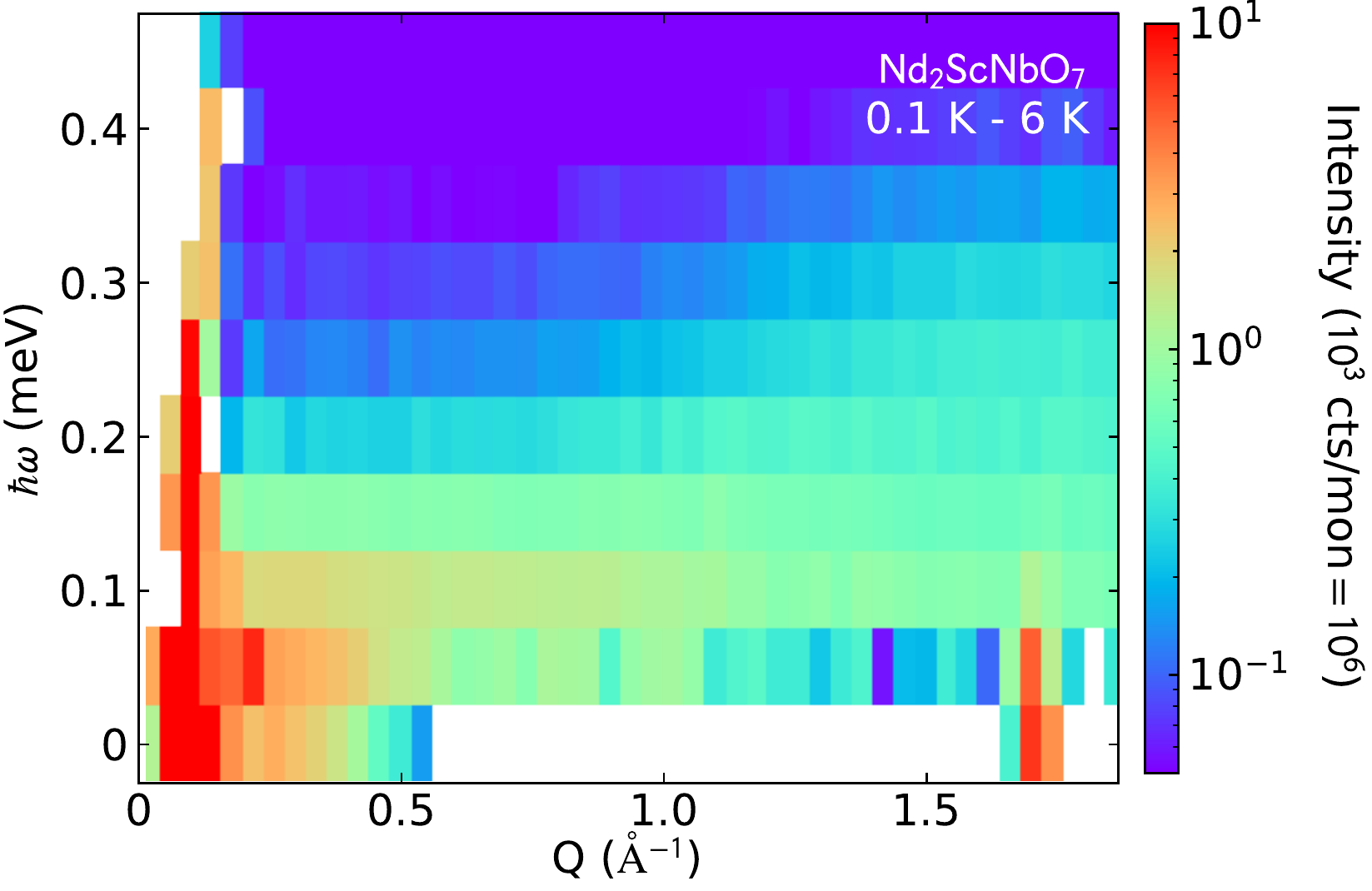}
	
	\caption{Neutron spectrum of $\rm Nd_2ScNbO_7$ at 0.1 K with 6 K background subtracted, showing an intense flat band at 0.1 meV and weaker scattering with a bandwidth of 0.4 meV. Note that the scattering comes down to $\hbar \omega = 0$ at low $Q$.}
	\label{flo:inelastic}
	
\end{figure}

In the MACS experiment, we measured the elastic scattering using a double-focusing configuration and $E_i = E_f = 5$~meV neutrons 
with beryllium filters in the incident and scattered beams at sample temperatures 0.1~K (below the phase transition) and 3~K (where the magnetic entropy is completely recovered). These data are shown in Fig \ref{flo:refinement}. There are clear temperature- dependent peaks which signal long-range magnetic order at low temperatures. The 5~meV neutron data have an energy resolution of 0.35 meV, which is roughly the bandwidth of the excitations (see Fig. \ref{flo:inelastic}), which makes it functionally an energy-integrated diffraction configuration for this sample.

We measured the inelastic spectrum with $E_f = 2.5$~meV neutrons (still with beryllium filters) for a full width at half maximum (FWHM)  energy resolution of 0.08 meV, covering energy transfers from $\hbar \omega = 0.0$~meV to $\hbar \omega = 0.5$~meV. The inelastic signal was measured at  $T=0.1$~K while a paramagnetic background was acquired at $T=6$~K. These data are shown in Fig. \ref{flo:inelastic}. There is an intense flat band excitation at 0.1 meV, very similar to the sister-compound  $\rm Nd_2Zr_2O_7$ \cite{Petit2016} (ref. \cite{mauws2019order} also observed the $\rm Nd_2ScNbO_7$ flat-band excitation and placed this band at 0.07 meV). This similarity suggests that whatever magnetic state exists in $\rm Nd_2Zr_2O_7$ also exists in $\rm Nd_2ScNbO_7$.

To determine the magnetic ordered structure, we performed a Rietveld refinement on the temperature-subtracted diffraction  data [Fig. \ref{flo:refinement}(b)] using the Fullprof software package \cite{Fullprof} and SARAh \cite{SARAh} to generate the irreducible representations. The only strongly-temperature dependent peaks are $(220)$ and $(311)$, but the fit nonetheless uniquely points to all-in-all-out (AIAO) order on the pyrochlore lattice with an ordered moment 1.121(9) $\rm \mu_B$. (There are perhaps small temperature-dependent features on the $(111)$ and $(200)$ peaks, but these are barely above the experimental error bar associated with the T-difference measurement, and no single Irrep was able to account for their intensity alongside the stronger magnetic  peaks.) The discovery of AIAO order is not surprising: nearly all Nd$^{3+}$ pyrochlores order with this structure: cf.  $\rm Nd_2Zr_2O_7$ \cite{Petit2016}, $\rm Nd_2Hf_2O_7$ \cite{Anand_2015},  $\rm Nd_2Sn_2O_7$ \cite{Bertin_2015}, and  $\rm Nd_2Ir_2O_7$ \cite{Guo_2016}.  

One thing that is unique about $\rm Nd_2ScNbO_7$ is the existence of temperature-dependent low-$Q$ diffuse scattering that appears to be magnetic (see Fig. \ref{flo:refinement}). Subsequent backscattering measurements showed this diffuse scattering to be real and elastic to within 0.4 $\mu$eV, with an onset at $T_N$ [Fig. \ref{flo:refinement}(d)]. This diffuse scattering carries 9(2)\% of the total integrated intensity of the (220) and (311) peaks (integrating $\int \frac{d^2 \sigma}{d \Omega d \omega} Q^2 d Q$ to account for the powder average). 
Despite the low-$Q$ diffuse scattering, the magnetic Bragg peaks are no wider than the nuclear Bragg peaks (the fitted correlation length from the (220) Bragg peak is $ > 2000$~\AA), indicating a very long magnetic correlation length.

Finally, we measured the temperature dependence of the (220) Bragg peak and found an order parameter curve consistent with a 3D Ising order [$\beta = 0.3265$ \cite{PELISSETTO2002}, Fig. \ref{flo:refinement}(c)] with a fitted critical temperature $T_N= 373(2)$~mK.

\subsubsection{Backscattering experiment}

\begin{figure}
	\centering\includegraphics[width=0.47\textwidth]{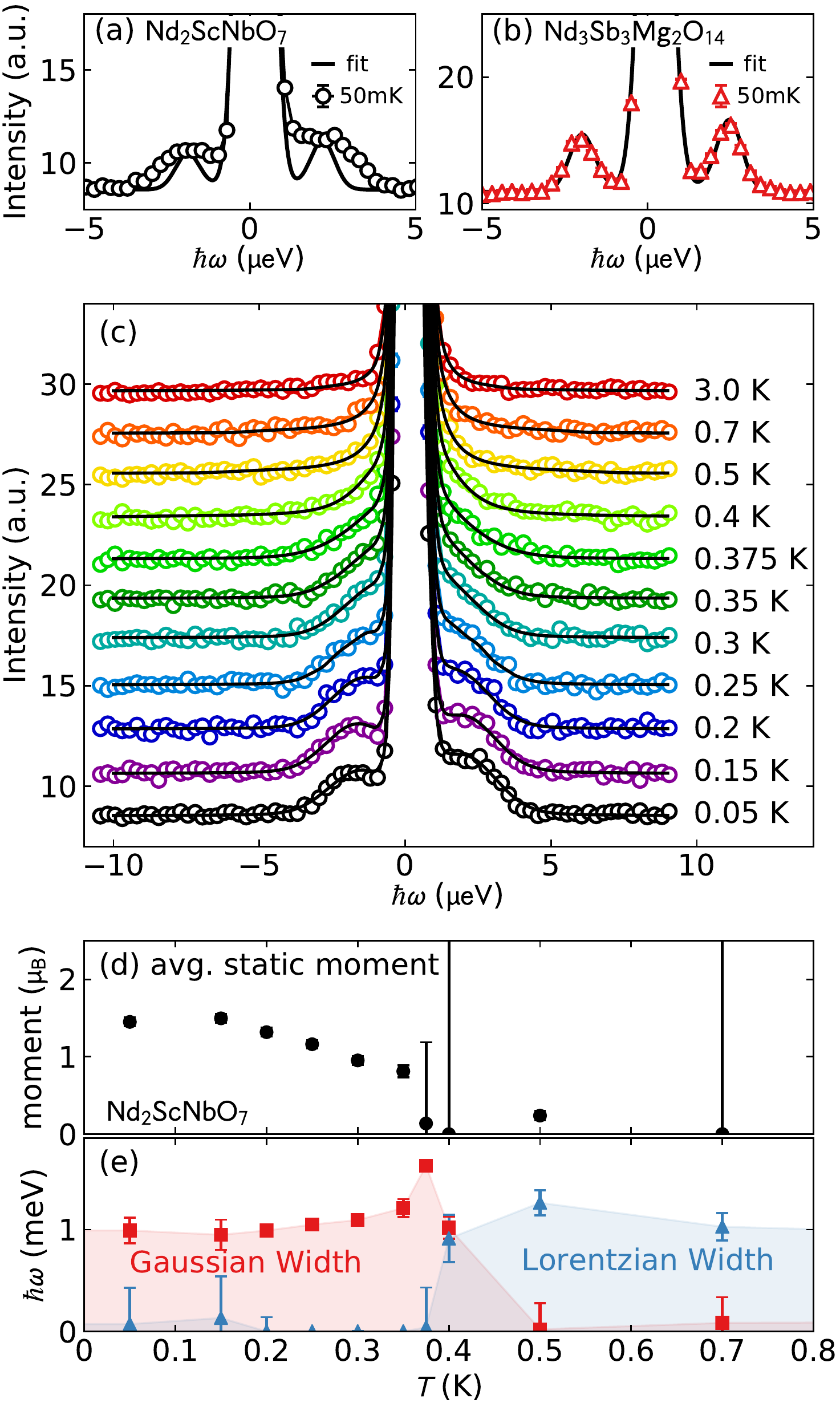}
	
	\caption{(a) $\rm Nd_2ScNbO_7$ Nuclear hyperfine neutron scattering on HFBS. (b) $\rm Nd_3Sb_3Mg_2O_{14}$ measured with the same instrument configuration. Comparison between (a) and (b) shows significant broadening of the nuclear hyperfine peaks in $\rm Nd_2ScNbO_7$, indicating a distribution of local static moments. (c) Hyperfine $\rm Nd_2ScNbO_7$ scattering as a function of temperature, with the fitted model. (d) Average static moment extracted from the model. (c)  Gaussian and Lorentzian widths to the nuclear hyperfine peaks. Below the transition temperature, Gaussian distributions dominate, but near and above the transition temperature Lorentzian broadening dominates. Error bars indicate one standard deviation.}
	\label{flo:HFBS_fit}
	
\end{figure}

In the HFBS experiment, we measured the excitation spectrum with a $\pm 11 {\rm \mu eV}$ bandwidth and an energy resolution FWHM of 0.8 ${\rm \mu eV}$. At this high resolution, the Nd nuclear hyperfine excitations become visible just off the central elastic peak (cf. refs. \cite{scheie2019homogenous,Chatterji2008,Bertin_2015}). These arise from a static electronic moment splitting the $2I+1=8$ degenerate nuclear spin levels via the nuclear hyperfine interaction. This gives rise to a $Q$-independent peak in the inelastic neutron scattering spectrum at an energy that corresponds to $\pm$ the level splitting.  \cite{heidemann1970}. When the nuclear hyperfine energy levels are known, the local static electronic moment can be calculated from the energies of these peaks \cite{Chatterji2008,scheie2019homogenous,Bertin_2015}. We measured the excitation spectrum as a function of temperature from 50 mK to 3 K. These data are shown in Fig. \ref{flo:HFBS_fit}.

As is immediately evident from Fig. \ref{flo:HFBS_fit}(a), the observed nuclear hyperfine excitations are significantly broader than the resolution width (which is defined by a vanadium scan---see Appendix \ref{app:Resolution}). We show this broadening to be intrinsic by comparing Fig. \ref{flo:HFBS_fit}(a) to the related compound $\rm Nd_3Sb_3Mg_2O_{14}$, shown in Fig. \ref{flo:HFBS_fit}(b) (from ref. \cite{scheie2019homogenous}). Both data sets were taken on the same instrument with the same sample environment and the same exchange gas pressure but $\rm Nd_2ScNbO_7$ has much broader nuclear hyperfine excitations: this shows the broadening is a property of  $\rm Nd_2ScNbO_7$. This broadening indicates either a static distribution of ordered moments or a short lifetime of the nuclear hyperfine levels that could result from electronic spin fluctuations.

The temperature dependence of these features is shown in Fig. \ref{flo:HFBS_fit}(c). At the lowest temperatures, two humps are visible on either side of the central elastic peak. As temperature increases, the hyperfine peaks shift into the central elastic peak and become Lorentzian-like tails.
We fit the HFBS data to a model including both Lorentzian and Gaussian broadening (described in Appendix \ref{app:HyperfineModel}), where the Gaussian is meant to account for a static moment distribution and the Lorentzian is meant to account for dynamic moments. The fits are shown as the black lines in Fig. \ref{flo:HFBS_fit}(c) and the widths as a function of temperature are shown in Fig. \ref{flo:HFBS_fit}(d). From these fits, we extract the mean ordered moment using the empirical relation between Nd nuclear hyperfine energies $\Delta E$ and static magnetic moment $\mu$ in ref. \cite{Chatterji2008},
\begin{equation}
\Delta E = \mu \times (1.25 \pm 0.04) {\rm \frac{\mu eV}{\mu_B}}.
\end{equation}
 At the lowest temperatures, the broadening appears to be mostly Gaussian with an average ordered moment of 1.47(6)~$\mu_B$ and a Gaussian spread of 0.8(1)~$\mu_B$ (see Fig. \ref{flo:HFBS_fit}(d)). Above $T=370$~mK the peak standard deviation overlaps with zero static moments and the broadening shifts to Lorentzian. This model is very rough, 
 but it is consistent with static moments for  $T<T_N$ and quasielastic scattering from fluctuating moments for $T>T_N$.

We also measured a high resolution elastic ($0 \pm 0.4 \> {\rm \mu eV}$) order parameter scan with the fixed-window mode, focusing on $Q=0.36(5) \>$\AA$^{-1}$ (Fig. \ref{flo:refinement}(d)). This showed the low-$Q$ diffuse scattering observed in Fig. \ref{flo:refinement}(b) to be associated with the magnetic order: its onset is at $T_N = 370(2)$~mK and its temperature dependence is consistent with the 3D Ising order parameter ($\beta = 0.3265$ \cite{PELISSETTO2002}), showing the low-$Q$ diffuse scattering to be magnetic in origin.  Allowing the critical exponent to vary in these order parameter fits, we find $\beta = 0.34(7)$ for the MACS data and $\beta=0.34(4)$ for the HFBS data---consistent with 3D Ising ($\beta = 0.3265$), 3D XY ($\beta = 0.3485$), and 3D Heisenberg ($\beta = 0.3689$) magnetic order \cite{PELISSETTO2002}; but the uncertainty is too high to pinpoint the anisiotropy of the exchange  interaction.

\section{Analysis}\label{sec:Analysis}

These experiments have shown $\rm Nd_2ScNbO_7$ to have easy-axis anisotropy and AIAO long-range magnetic order.  These features are similar to many other Nd$^{3+}$ pyrochlores.  Combining the Neel temperatures measured from heat capacity and the two order parameter curves, we find a mean Neel temperature $T_N = 371(2)$~mK.

These experiments also reveal some unusual features: (A) an unusual temperature dependence in heat capacity, (B) a discrepancy between ordered moments obtained from different experimental methods, and (C) a distribution in static magnetic moments evidenced by broadened hyperfine excitations.
Considered together, they indicate an unconventional magnetic ground state with strong quantum fluctuations.

\subsection{Density of states in heat capacity}

\begin{figure*}
	\centering\includegraphics[width=\textwidth]{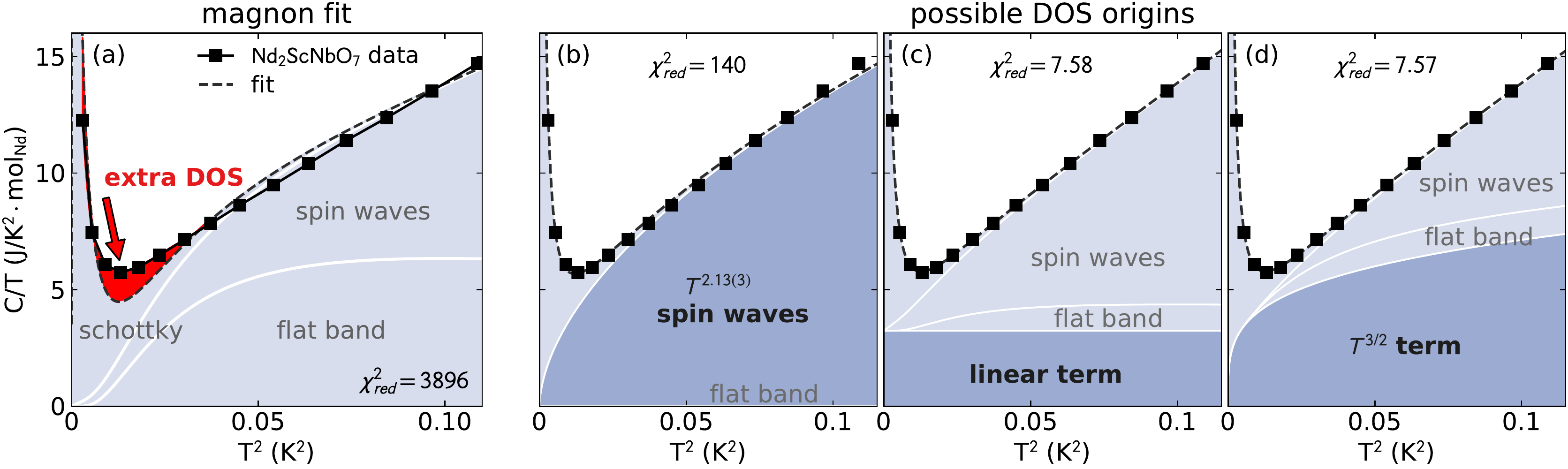}
	
	\caption{Low temperature heat capacity of $\rm Nd_2ScNbO_7$ compared to various models. (a) shows a fitted model based on an 0.07 meV flat band spin wave excitation plus linear dispersive spin waves and a schottky anomaly. It fails to properly account for the density of states at low T, indicating some exotic behavior. (b) shows the same fitted model but the heat capacity from spin wave excitations are raised to an arbitrary fitted power. (c) shows a fitted model which includes a linear term in specific heat. This linear term model fits the data two orders of magnitude better than the model in (b). (d) shows a fit including a spinon $T^{3/2}$ contribution as predicted by ref. \cite{Liu_2019}.}
	\label{flo:LowT_DOS}
\end{figure*}

The first unusual feature is in the temperature dependence of the magnetic heat capacity below $T_N$. 
 Our initial phenomenological fit showed that the electronic heat capacity does not follow the expected $T^3$ behavior of linear dispersive modes in three dimensions, but a power law closer to $T^{2.2}$. 
The peculiarity becomes more apparent when we plot $C/T$ vs $T^2$, as in Fig. \ref{flo:LowT_DOS}. This shows a very nearly linear trend, but with a positive offset in $C/T$ (i.e., the heat capacity seems to follow $C = \gamma T + \alpha T^3$ with $\gamma > 0$). This is opposite to expectations for gapped Bosonic modes: in general, heat capacity from a gapped bosonic spectrum produces a \textit{negative} offset in $C/T$ \cite{scheie2019homogenous}. 

Several things can cause non-$T^3$ heat capacity in a 3D antiferromagnetic insulator: (i) a spin wave spectrum that cannot be approximated by gapped linear dispersive modes (such as in $\rm Gd_2Sn_2O_7$ \cite{Quilliam_2007,Maestro_2007}). (ii)  glassy spin disorder which produces a non-integer power law in $T$ \cite{Thomson_1981,SHEIKH1989}. (iii) Exotic excitations which can produce positive  non-$T^3$ terms in heat capacity \cite{Yamashita2008,Liu_2019}. We test each of these scenarios by building an appropriate model and comparing it with the data. (Details of our fitting methods are in Appendix \ref{app:HCmodelFits}.)

We test (i) by building a phenomenological spin-wave model. Having observed the low-energy flat mode (Fig. \ref{flo:inelastic}), we infer the following magnetic  Einstein term in the specific heat: 
\begin{equation}
C_{flat} = k_B \big( \frac{\hbar \omega_0}{k_B T} \big)^2 \frac{ e^{\hbar \omega_0/ k_B T}}{ ( e^{\hbar \omega_0/ k_B T}-1)^2}
\label{eq:EinsteinMode}
\end{equation}
with $\hbar \omega_0 =0.07$~meV, and we insert it into the following model:
\begin{equation}
C = A \times C_{flat} + B \times C_{ld}(\Delta, c) + C_{schottky}(\mu)
\label{eq:MagnonModel}
\end{equation}
where $C_{ld}(\Delta, c)$ is the heat capacity of a gapped linear dispersive spectrum of the form $\eta (q) = \sqrt{\Delta^2 + (cq)^2}$ (calculated as in ref. \cite{scheie2019homogenous}), and $C_{schottky}(\mu)$ is a Schottky anomaly. $A$, $B$, $\Delta$, $c$, and $\mu$ are fitted parameters. The best fit is shown in Fig. \ref{flo:LowT_DOS}(a) and the best fit parameters are given in Table \ref{tab:Magnon+FlatBand_BestFit}. This fit underestimates heat capacity around 0.1 K and does not match the temperature-dependence at higher temperatures. This indicates a low-energy density of states (DOS) below 0.07 meV not accounted for by this model. There are clearly some significant contributions below the flat band that are not linear dispersive spin waves.

%

We test (ii) by re-doing the fit, this time allowing the dispersive spin wave contribution to take on an arbitrary power in temperature:
\begin{equation}
C = A \times C_{flat} + B \> T^n + C_{schottky}(\mu)
\label{eq:PowerLawModel}
\end{equation}
where  $A$, $B$, $n$, and $\mu$ are fitted parameters. The result of this fit is shown in Fig. \ref{flo:LowT_DOS}(b) and the best fit parameters are given in Table \ref{tab:PowerLaw+FlatBand_BestFit}. The fit is much improved, but the model still does not match the temperature dependence above 0.05~K$^2$.

We test (iii) by trying two models. First, we add a linear term in the heat capacity to account for fermionic quasiparticles \cite{Yamashita2008,Kittel}:
\begin{equation}
C = A \times C_{flat} + B \> T^3 + \gamma T + C_{schottky}(\mu)
\label{eq:LinearTermModel}
\end{equation}
where  $A$, $B$, $\gamma$, and $\mu$ are fitted parameters. The result of this fit are shown in Fig. \ref{flo:LowT_DOS}(c) and Table \ref{tab:Magon+LinearTerm_BestFit}. The match is very good: there are virtually no deviations from the experimental data points, and the reduced $\chi^2$ is improved over the power-law fit by nearly two orders of magnitude.

Second, we consider a $T^{3/2}$ term to account  for condensed spin-liquid spinons as in ref. \cite{Liu_2019}:
\begin{equation}
C = A \> C_{flat} + C_{ld}(\Delta, c) + B \> T^{3/2} + C_{schottky}(\mu)
\label{eq:T3/2model}
\end{equation}
where $A$, $\Delta$, $c$, $B$, and $\mu$ are fitted parameters. The result of this fit are shown in Fig. \ref{flo:LowT_DOS}(d) and Table \ref{tab:Magon+T3/2_BestFit}. The match is again very good. This model included one additional fitting parameter as compared to the others, but the reduced $\chi^2$ shows that the match is nearly equivalent to the linear term model. 

As a side-note, these fits better constrain the ordered moment from the Schottky anomaly. 
If we take the three fits which most accurately model the electronic specific heat (the linear term fit and the $T^{3/2}$ fit) and the initial phenomenological fit, the weighted mean ordered moment is $\mu=1.171(4) {\rm \mu_B}$ with a standard deviation of $0.03 {\rm \mu_B}$. Here we take the standard deviation as uncertainty reflecting the uncertainty in which model is correct, and so the ordered moment from the nuclear Schottky anomaly is $\mu=1.17(3) {\rm \mu_B}$.

\subsubsection*{Comparison to other Nd$^{3+}$ pyrochlores}

\begin{figure}
	\centering\includegraphics[width=0.47\textwidth]{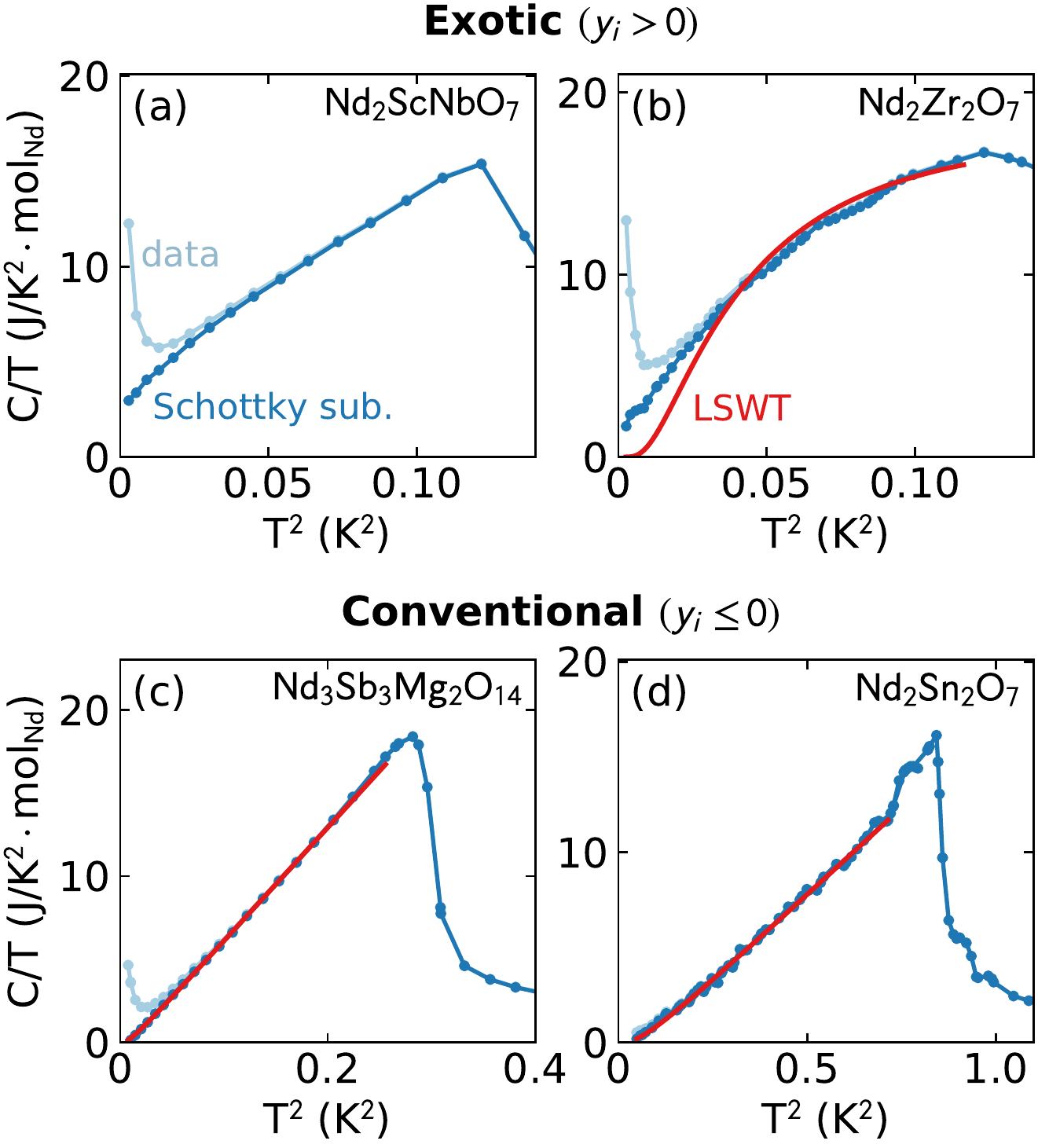}
	
	\caption{Heat capacity of $\rm Nd_2ScNbO_7$ plotted in $C/T$ vs $T^2$ (a) compared to three other Nd$^{3+}$ pyrochlore-based compounds: $\rm Nd_2Zr_2O_7$ (b) \cite{Blote1969}, $\rm Nd_3Sb_3Mg_2O_{14}$ (c) \cite{MyPaper}, and $\rm Nd_2Sn_2O_7$ (d) \cite{Blote1969}. (c) and (d) show conventional spin-wave heat capacity, while (a) and (b) show excess heat capacity at low temperatures, even compared to heat capacity computed directly from the spin wave spectrum (b).  Note that the nuclear hyperfine contribution to the specific heat was subtracted from all data, including $\rm Nd_2Sn_2O_7$ shown in (d).}
	\label{flo:HC_comparison}
\end{figure}

This exercise in fitting shows that---given reasonable assumptions about collective bosonic excitations---the density of states in $\rm Nd_2ScNbO_7$ cannot be accounted for by such quasiparticles alone. 
This conclusion is bolstered by comparison to the heat capacity of $\rm Nd_2Zr_2O_7$, which is also quite unusual with an apparent low-temperature $T^2$ dependence \cite{Blote1969,Xu_2019_Anisotropic}. Fortunately, the $\rm Nd_2Zr_2O_7$ spin wave spectrum has been measured and modeled in detail \cite{Xu_2019_Anisotropic}, so it is possible to directly calculate the magnon heat capacity in the low $T$ limit (details are in Appendix \ref{app:HeatCapacityFromSpinWaves}). The results are depicted in Fig. \ref{flo:HC_comparison}(b), and show that the measured specific heat is much higher than the calculated magnon specific heat at low temperatures where higher order effects associated with the phase transition (damping and softening of collective modes) can be neglected. This indicates an anomalous low-energy DOS that cannot be accounted for through conventional spin wave theory ---just like $\rm Nd_2ScNbO_7$. 

Now this unusual heat capacity is not present in all Nd$^{3+}$ pyrochlores.  Fig. \ref{flo:HC_comparison}(c)-(d) show two Nd$^{3+}$ pyrochlore-based materials with conventional magnon behavior: $T^3$ heat capacity with $- \gamma$ offsets.
 \ref{flo:HC_comparison}(c) shows data from the Nd$^{3+}$ pyrochlore derivative $\rm Nd_3Sb_3Mg_2O_{14}$ (a kagome compound with AIAO order and easy-axis anisotropy) which has heat capacity exactly matching gapped linear dispersive magnon modes \cite{scheie2019homogenous}. Fig. \ref{flo:HC_comparison}(d) shows data from pyrochlore $\rm Nd_2Sn_2O_7$ \cite{Blote1969} (another Nd$^{3+}$ pyrochlore with AIAO order and easy-axis anisotropy \cite{Bertin_2015}), is beautifully consistent with gapped dispersive magnons indicated by the solid red lines (further details are in Appendix \ref{app:HeatCapacityFromSpinWaves}).
 Thus, not all Nd$^{3+}$ pyrochlores have the extra DOS observed in $\rm Nd_2ScNbO_7$ and $\rm Nd_2Zr_2O_7$. 
Specifically, only the specific heats of the two materials with gapped excitation spectra are well accounted for by thermal excitation of bosonic collective modes.

\subsection{Static moment discrepancy}

Another unusual feature of $\rm Nd_2ScNbO_7$  is the apparent discrepancy in the measured static moment. Specifically under the assumption of a static spin structure, three different measurements (nuclear Schottky anomaly, diffraction refinement, and hyperfine excitations) yield different values for the ordered moment size, shown in Table \ref{tab:orderedmoment}. All are less than half of the single site saturation magnetization inferred from magnetization measurements on a powder sample. 
The Schottky anomaly and the neutron diffraction refinement values agree to within uncertainty, but there is an apparent discrepancy with the value derived from the nuclear hyperfine excitations.

\begin{table}[H]
	\caption{Low temperature ordered electronic moment $\mu$ of $\rm{Nd_2ScNbO_{7}}$ measured by nuclear Schottky anomaly, neutron diffraction, and hyperfine excitations. The bottom row describes what quantity each technique measures.}
	\begin{ruledtabular}
		\begin{tabular}{ccc}
			Nuclear & Neutron & Hyperfine \tabularnewline
			Schottky & diffraction & excitations \tabularnewline
			\hline 
			1.17(3) $\rm \mu_B$  & 1.121(9) $\rm \mu_B$ & $1.47(6)  ~ \mu_B$  \tabularnewline[0.15cm] 
			\textit{Local RMS $\mu$} & \textit{Mean LRO $\mu$} & \textit{Mean local $\mu$} 
	\end{tabular}\end{ruledtabular}
	\label{tab:orderedmoment}
\end{table}

To interpret these differences we must consider what quantity each technique measures. The neutron diffraction measurement is a spatially-averaging probe: it gives only the static moment (on a time scale set by the energy resolution) that participates in a long-range-ordered (LRO) state. Meanwhile, the measurements based on the nuclear spin (Schottky anomaly and hyperfine excitations) are local probes: they give the value of the electronic ordered moment without reference to spatial correlations. There are important differences between the hyperfine excitation and nuclear Schottky probes too: the high-temperature tail of a nuclear Schottky anomaly measures the root-mean-squared (RMS) local static moment (at intermediate temperatures it measures a value between the mean and RMS, see Appendix \ref{app:NuclearSchottky}) while the hyperfine excitation spectrum is a measure of the probability distribution function for the magnitude of the local moment. 

It is possible for the local static moment to be larger than the LRO moment if there is static spin disorder \cite{HTO_nuclearmagnetism}. This is what one expects for the moment fragmented state on the pyrochlore lattice: in that case, the local static moment would be twice the LRO moment \cite{Brooks-Bartlett_2014,Petit2016}. However, this is not what we observe: the local moment is only  $\frac{1.47(6)}{1.121(9)} = 1.32(5)$ times larger than the LRO moment, and both are much smaller than the sublattice saturation magnetization inferred from the powder magnetization measurements. Thus $\rm Nd_2ScNbO_7$ does not have a moment fragmented ground state (in the sense of refs. \cite{Brooks-Bartlett_2014,Petit2016,Paddison2016,Lefrancois2017fragmented} referring to a three-in-one-out crystallized lattice of magnetic monopoles).
The fragmentation theory notwithstanding, static spin disorder (from, for example, a distribution of moment sizes) may explain why the static moment inferred from the nuclear spin polarization is greater than the moment inferred from magnetic neutron diffraction.

The most curious discrepancy is in the difference between the moment inferred from the nuclear Schottky anomaly and the hyperfine spectrum. If all moments are uniform size, these values should be the same (cf. ref. \cite{scheie2019homogenous}). With spin disorder, the nuclear Schottky moment should exceed the mean value from hyperfine excitations (by definition, RMS $\geq$ mean). However, we see the opposite: the RMS nuclear Schottky value is less than the mean nuclear hyperfine value. The same three measurements were performed using the exact same equipment on the related Nd magnet $\rm Nd_3Sb_3Mg_2O_{14}$ \cite{scheie2019homogenous} (which has no exotic features in its ground state) and those three measurements agree beautifully---so the measurement technique, calibration, and analysis methods seem to be in order.
Rather, this discrepancy and the reduction seen in all these moment measurements relative to the sublattice saturation magnetization suggests that some fraction of the Nd electronic moments is fluctuating.

\begin{figure}
	\centering\includegraphics[width=0.47\textwidth]{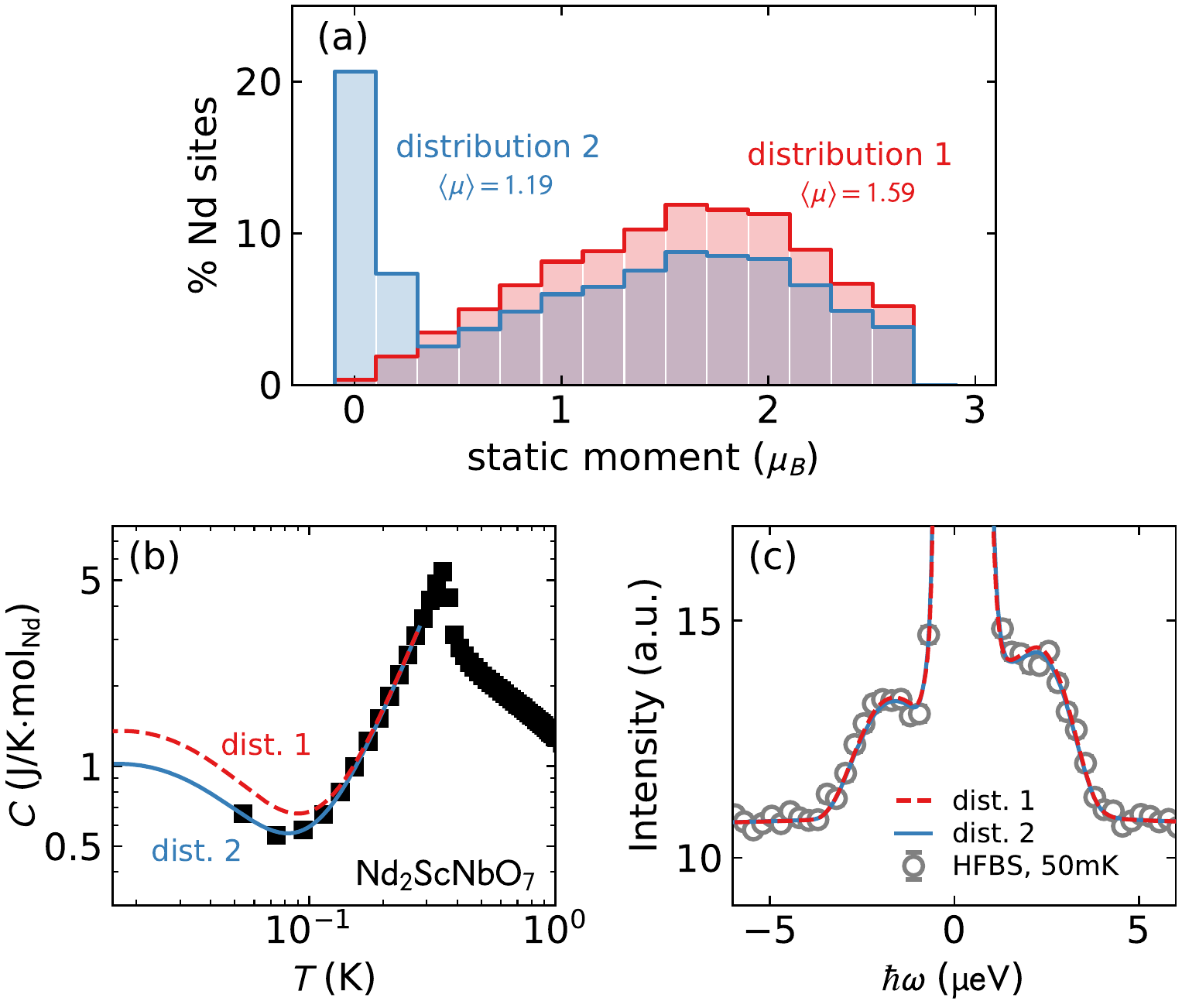}
	
	\caption{(a) Two possible distributions of static moments in $\rm Nd_2ScNbO_7$. Distribution 1 (red) was modeled after the positive-energy transfer hyperfine spectrum. Distribution 2 (blue) is the same distribution, but with 1/4 of the moments having zero or nearly zero static moments. (b) Calculated Schottky heat capacity from the two distributions; distribution 2 matches the data very well. (c) Hyperfine spectrum calculated from distributions 1 and 2, normalized to match the hyperfine scattering. They are identical: the zero-moments from model 2 are hidden in the elastic peak.}
	\label{flo:MomentHistogram}
\end{figure}

If the moment measured from the hyperfine spectrum is larger than the moment from the Schottky anomaly, this indicates that the hyperfine measurement missed some sites with very small moments.
A limitation of fitting hyperfine scattering is that signals from vanishing static magnetism are hidden in the elastic scattering, and the normalization is not precise enough to identify the missing spectral weight. Thus, if the distribution of Nd moments is not Gaussian (as our model above assumed) and a significant fraction of Nd sites have vanishing static moments (small enough to be hidden in the elastic channel), the fitted mean moment would be higher than the moment revealed by a Schottky anomaly fit. 
This is demonstrated in Fig. \ref{flo:MomentHistogram}, where two different distributions reproduce the hyperfine spectrum but have different mean moments  and nuclear Schottky signals. Distribution 1 ($\langle \mu \rangle = 1.59 \> \mu_B$, $\mu_{RMS} = 1.71 \> \mu_B$) follows the profile of the positive energy transfer hyperfine spectrum, but distribution 2 ($\langle \mu \rangle = 1.19 \> \mu_B$, $\mu_{RMS} = 1.46 \> \mu_B$) has 1/4 of moments clustered around zero. Distribution 2 is consistent with both neutron and heat capacity data. (Note that the RMS moment is greater than the fit above: this is because at low temperatures the fitted moment deviates from the RMS --- see Appendix \ref{app:NuclearSchottky}.)
Under this interpretation, the static moment discrepancy shows some fraction of Nd sites have fluctuating moments to the lowest temperatures, such that the static moment on some sites is nearly zero.

\subsection{Ordered moment distribution}

A third question raised by these results is why there is a distribution in ordered moment size indicated by the backscattering measurements. Although we cannot say for certain, it is possible that this distribution is due to variations in the local crystal fields of Nd from the Nb-Sc B-site disorder. To demonstrate this, we simulated the crystal field environment using a point charge model.

\begin{figure*}
	\centering\includegraphics[width=\textwidth]{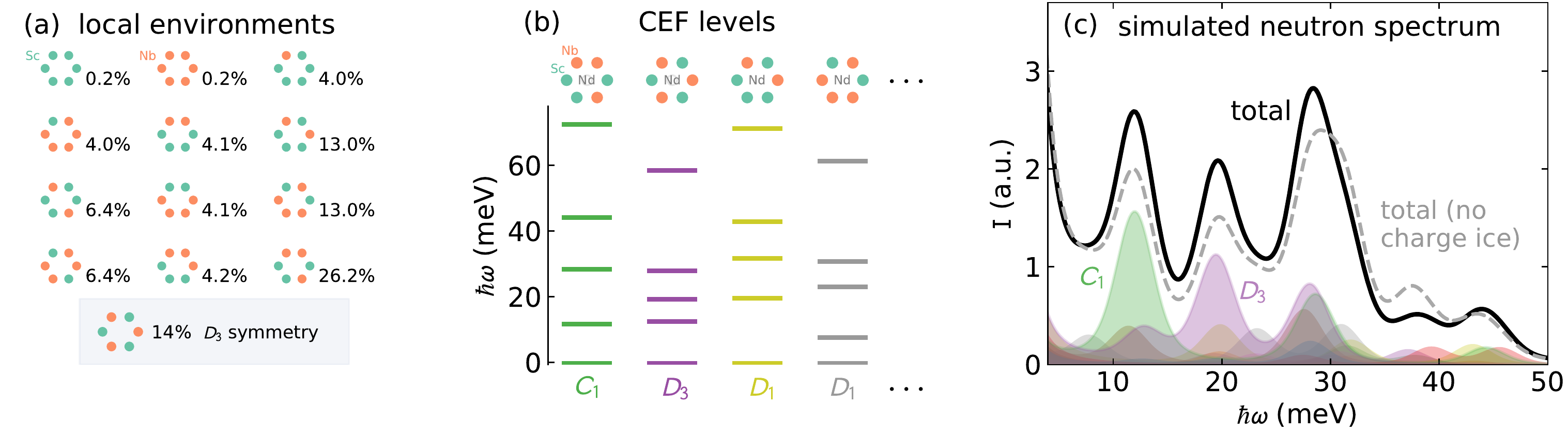}
	
	\caption{Simulated CEF spectrum for $\rm Nd_2ScNbO_7$. (a) Distribution of symmetry-equivalent local Nd environments if 14\% are $D_3$. (b) CEF levels of the four most common local environments ($C_1$: 26.2\%, $D_3$: 14\%, and both $D_1$: 13\%) calculated from a point-charge model assuming modest oxygen displacement. (c) Total simulated neutron spectrum from all CEF environments, showing that observable peaks are from more than just $D_3$ environments. The light grey dashed line shows the total scattering assuming no charge ice correlations (totally random Nd and Sc placement).}
	\label{flo:CEF_simulation}
	
\end{figure*}

Measurements by Mauws et al \cite{mauws2019order} show a distribution of crystal electric field (CEF) excitations in $\rm Nd_2ScNbO_7$. Based on the relative peak weights, they estimate that 14\% of the sites have $D_3$ symmetry. Assuming this to be true, we performed Monte Carlo simulations of Nb$^{5+}$ Sc$^{3+}$  charge ice at finite temperature to determine the relative frequency of the other symmetry-unique sites (see Appendix \ref{app:MonteCarlo}). Although there are $2^6 = 64$ possible Nb-Sc arrangements around a Nd ion, there are only 13 symmetry-unique arrangements.  When 14\% of Nd$^{3+}$ sites have $D_3$ symmetry, the percentages of other symmetry-unique CEF environments are shown in Fig. \ref{flo:CEF_simulation}(a).

We then calculated the CEF Hamiltonians of the various CEF states using the point charge approximation and PyCrystalField \cite{PyCrystalField} (see Appendix \ref{app:MonteCarlo} for details), and the levels of the four most probable states are plotted in Fig. \ref{flo:CEF_simulation}(b). Because Nd$^{3+}$ is a Kramers ion, every environment yields five doublets no matter how low the symmetry. We then simulated the neutron spectrum by summing over all CEF states with the appropriate weights [Fig. \ref{flo:CEF_simulation}(c)] and found a spectrum which resembles what was measured in ref. \cite{mauws2019order}. However, we find that many of the strongest peaks are not from $D_3$ sites. If instead we assume a completely random Nb and Sc distribution with no charge-ice correlations [shown by the grey line in Fig. \ref{flo:CEF_simulation}(c)] the calculated neutron spectrum is only moderately changed---and possibly corresponds to the data in ref. \cite{mauws2019order} even better. These simulations cast doubt on the methodology behind the 14\% estimate in ref. \cite{mauws2019order}.

In any case, assuming that these calculations give at least an approximate picture of the distribution of CEF states, we find that the maximum static moments for these sites have a mean value $\mu=1.645 \> \mu_B$ with a standard deviation $1.117 \> \mu_B$. This is very close to the observed $1.47(6) \pm 0.8(1) ~ \mu_B$ from the hyperfine excitations. Therefore the ordered moment distribution might be partially caused by disordered crystal field environments.

It should be noted that the precise values of the calculated static moments are dependent on the details of the point-charge simulations. Different assumptions about oxygen displacement produced a higher mean $\mu$ and a lower standard deviation. Furthermore, the hyperfine spectrum simulated directly from the CEF model shows more discrete peaks than the observed hyperfine spectrum (see Appendix \ref{app:MonteCarlo}), possibly because of the simplistic point-charge model and collective effects which affect the low temperature static moments. Therefore, these calculations demonstrate the possibility (but do not prove) that the distribution of ordered moments is caused by B-site disorder.

With this moment size distribution, one would expect to see some diffuse elastic scattering indicating a superposition of random spin orientation with long-range AIAO order: the disordered moment estimated from nuclear hyperfine excitations is $\frac{(0.8(1)  \> \mu_B)^2}{(1.47(6)  \> \mu_B)^2} = 26(7)$\%. This may explain the observed low-Q diffuse elastic signal in Fig. \ref{flo:refinement}. However, the low-$Q$ scattering decreases with $Q$ much more dramatically than the Nd form factor, indicating correlated spin disorder rather than random disorder.

\section{Discussion}

The results from heat capacity and local moment measurements indicate a strongly fluctuating magnetic state with an unusual DOS which seems to be common among several Nd$^{3+}$ pyrochlores.
It is clear that conventional magnons cannot account for the DOS in $\rm Nd_2ScNbO_7$ and  $\rm Nd_2Zr_2O_7$, but it is not clear what kind of excitations they are.
In principle, the apparent linear offset could be from a \textit{very} low-energy bosonic flat mode. If the energy of the mode is $\sim 5 ~ {\rm \mu eV}$, the heat capacity signal would look very much like a linear term. However, such a strong low-energy mode would have been visible in the neutron backscattering experiment, so we can constrain any such mode to have an energy greater than  $ 11  ~ {\rm \mu eV}$ or less than $0.4  ~ {\rm \mu eV}$. Furthermore, such a low-energy mode is not predicted in the spin-Hamiltonian of $\rm Nd_2Zr_2O_7$ \cite{Benton_2016,Xu_2019_Anisotropic}, rendering any such mode to be beyond LSWT magnons.
We present two possible explanations for the low-energy DOS: fermionic quasiparticles and local spin degrees of freedom.

Linear $C/T$ offsets are ordinarily signals of fermionic quasiparticles, and the $\gamma$ term (called a Sommerfeld coefficient) is a measure of the density of states at the chemical potential, which is proportional to the effective mass of the fermions. In $\rm Nd_2ScNbO_7$ we measure $\gamma = 3.08(4) \> {\rm \frac{J}{mol \> K^2}}$. This is three orders of magnitude higher than Sommerfeld coefficients for most metals ($\gamma_{\rm Ag} = 6.1\times 10^{-4} {\rm \frac{J}{mol \> K^2}}$, $\gamma_{\rm Nb} = 8.7\times 10^{-3} {\rm \frac{J}{mol \> K^2}}$ \cite{Kittel}), and is of the order of heavy-fermions systems such as YbBiPt, which has $\gamma \approx 8 {\rm \frac{J}{mol \> K^2}}$ \cite{Fisk1991}. 
This $\gamma$ could possibly be the signal of fermionic spinons of a quantum spin liquid \cite{Yamashita2008}, which is plausible given the theoretical proximity of  $\rm Nd_2Zr_2O_7$ to a $U(1)$ quantum spin liquid \cite{Benton_2016}.
Alternatively, it could be that the heat capacity signal is from $C \propto T^{3/2}$ spinons predicted for a condensed AIAO pyrochlore spin liquid phase \cite{Liu_2019}. In either case, this correspondence suggests that the mystery DOS in $\rm Nd_2ScNbO_7$ and  $\rm Nd_2Zr_2O_7$ are spinons of a proximate quantum spin liquid.

Under this interpretation, the distribution of moments may be due to a moment-modulated spin fragmented state, as proposed for $\rm Ho_3Sb_3Mg_2O_{14}$ \cite{dun2019quantum}. In this phase, dipolar interactions induce a quantum moment fragmented state where certain spins fluctuate more than others, leading to a distribution of static moments. Although the moment size for Nd is smaller than Ho, similar physics from asymmetric exchange may be relevant. If so, the specific fraction 1/4 of spins having nearly zero moment, which is consistent with the nuclear hyperfine spectrum, could indicate the "one-out" spin fluctuates while the "three-in" spins are static.

Alternatively, it could be that the mystery DOS is associated with local spin degrees of freedom induced by the Nd-Sc disorder.
It is known that specific heat in disordered spin glasses can have $T$-linear or $T^{3/2}$ specific heat \cite{Binder1986,Thomson_1981}. If such a heat capacity signal were superimposed upon magnon specific heat, the result could look very much like the fits in Fig. \ref{flo:HC_comparison}. 
Under this interpretation, the CEF and magnetic exchange disorder from the Nd-Sc disorder create local soft spin degrees of freedom within a long-range ordered state. This may explain the low-energy low-$Q$ scattering in Fig. \ref{flo:inelastic}.

This disorder hypothesis also naturally explains how a fraction of the spins could have no static moment. It has been shown theoretically that random bond disorder 
on the pyrochlore lattice can produce random singlet formation between spins \cite{Uematsu_2019}, or even long-range entangled states \cite{Savary2017}. Given how close $\rm Nd_2Zr_2O_7$ is to a $U1$ quantum spin liquid phase \cite{Benton_2016} it is possible that $\rm Nd_2ScNbO_7$ is also close enough that bond and anisotropy disorder causes certain site pairs to be locally within that phase and form singlets, such that there is no static moment on those sites and low-energy singlet-triplet excitations appear in the DOS.

 If this interpretation is correct, $\rm Nd_2ScNbO_7$ hosts the unusual combination of long-range magnetic order, singlet formation, glassy local excitations, and a sharp phase transition. This would imply that the disorder is weak enough that the long-range order is preserved (with a resolution-limited correlation length) though the exchange and single ion anisotropy disorder produces local soft modes for the spins (some of which pair to form singlets) and diffuse low $Q$ scattering.

Assuming this to be true, the exotic behavior of the sister-compound $\rm Nd_2Zr_2O_7$ may also be associated with  disorder-induced local spin degrees of freedom (perhaps through oxygen deficiency and "stuffing" problems which plague other pyrochlores \cite{Arpino2017,GHASEMI2018,Ruminy2016}). This disorder hypothesis may explain the dramatic sample-dependence reported in ref. \cite{Blote1969} and why different studies with different samples report conflicting values of the ordered moment \cite{Lhotel_2015,Petit2016,Xu_2015}.

Whether the mysterious magnetic  DOS is from fractionalized quasiparticles or local excitations, the resulting ground state is clearly dominated by strong quantum fluctuations.
Comparison between nuclear hyperfine excitations and specific heat data indicates there are sites where the electronic moment continues to fluctuate faster than the nuclear spin relaxation time so that no hyperfine level splitting is generated. Thus, quantum fluctuations persist to the lowest temperatures. 
An apparent suppression of ordering was also documented for 
 $\rm Gd_2Sn_2O_7$, and it was proposed that  quantum fluctuations in the electronic moment suppress a nuclear Schottky anomaly \cite{Bertin2002}.
The quantum fluctuations in $\rm Nd_2ScNbO_7$ could be from an exchange-disordered proximate spin liquid \cite{Kawamura_2019}, or localized excitations from a random singlet state. 
Occam's razor leads us to prefer the localized DOS hypothesis over the exotic quasiparticle hypothesis, but transport experiments \cite{Chen2019transport} are needed to determine whether the excitations are localized spin degrees of freedom or delocalized quasiparticles which can transport energy. 

 
These results have strong implications for other Nd pyrochlores. The many similarities between $\rm Nd_2ScNbO_7$ and  $\rm Nd_2Zr_2O_7$ suggests a common magnetic state, and we expect these same features to be present in $\rm Nd_2Hf_2O_7$ because of the close correspondence with  $\rm Nd_2Zr_2O_7$.  The magnetic ground state is strongly influenced by collective quantum physics: The combination of frustration and disorder in $\rm Nd_2ScNbO_7$  produces a state which fluctuates to the lowest temperatures and mimics spin-liquid behavior---but may not be the long-sought many-body entangled spin liquid state.

\section{Conclusion}

We have shown the Nd$^{3+}$ pyrochlore $\rm Nd_2ScNbO_7$ to have an easy-axis moment with AIAO magnetic order below $T_N=371(2)$ mK. Susceptibility and magnetization indicate an easy-axis moment, heat capacity shows a fully-ordered system with $R \ln2$ entropy, and elastic neutron scattering shows AIAO order.
Inelastic scattering shows a flat-band excitation similar to $\rm Nd_2Zr_2O_7$, and all order parameter curves are consistent with 3D Ising magnetic order. Nuclear hyperfine excitations reveal a distribution of static electronic moments below $T_N$, which we suggest is due to B-site disorder influencing the Nd$^{3+}$ crystal electric field states.

Analysis of our results has revealed two unconventional behaviors in $\rm Nd_2ScNbO_7$:
First, we used heat capacity to show an anomalous density of states in both $\rm Nd_2ScNbO_7$ and $\rm Nd_2Zr_2O_7$, that is inconsistent with conventional magnons. These excitations give a heat capacity signal consistent with fractionalized spinons of a proximate spin liquid phase, or localized spin excitations from a random-singlet phase that coexists with long range order.
Second, by comparing the local and long-range ordered moments, we rule out the possibility of classical moment fragmentation (in the sense of refs. \cite{Brooks-Bartlett_2014,Petit2016,Hermele_2014}) in $\rm Nd_2ScNbO_7$. This comparison of local ordered moments reveals that a fraction of Nd sites have vanishing static moments in the low $T$ limit, as in a random singlet state or an exotic moment-modulated quantum fragmented state.
Taken together, these results indicate a spin system dominated by quantum fluctuations that cannot be described in terms of conventional magnons in an otherwise long range ordered state. 
Disorder however, plays a significant role so that experiments probing the degree of spatial coherence of the low energy excitations will be needed to properly classify the unusual quantum state of matter in Nd-based pyrochlore magnets that we have drawn attention to in this paper. 


\subsection*{Acknowledgments}
Thanks to Christian Balz, Sayak Dasgupta, G\'abor Hal\'asz, Mark Lumsden, and Steven Nagler for helpful discussions.
This research used resources at the Spallation Neutron Source, a DOE Office of Science User Facility operated by the Oak Ridge National Laboratory.
Initial stages were supported as part of the Institute for Quantum Matter, an Energy Frontier Research Center funded by the U.S. Department of Energy, Office of Science, Basic Energy Sciences under Award No. DE-SC0019331.
AS and CB were supported by the Gordon and Betty Moore foundation under the EPIQS program GBMF4532.
The dilution refrigerator used for heat capacity measurements was funded by the NSF through DMR-0821005.
Use of the NCNR facility was supported in part by the NSF through DMR-1508249. Thanks to Jianhui Xi for sharing his SpinW code.

%
\quad


\appendix

\renewcommand{\thefigure}{A\arabic{figure}}
\renewcommand{\thetable}{A\arabic{table}}

\setcounter{figure}{0}

\section{Backscattering experiment and analysis} 

\subsection{HFBS resolution function}\label{app:Resolution}

\begin{figure}
	\centering\includegraphics[width=0.46\textwidth]{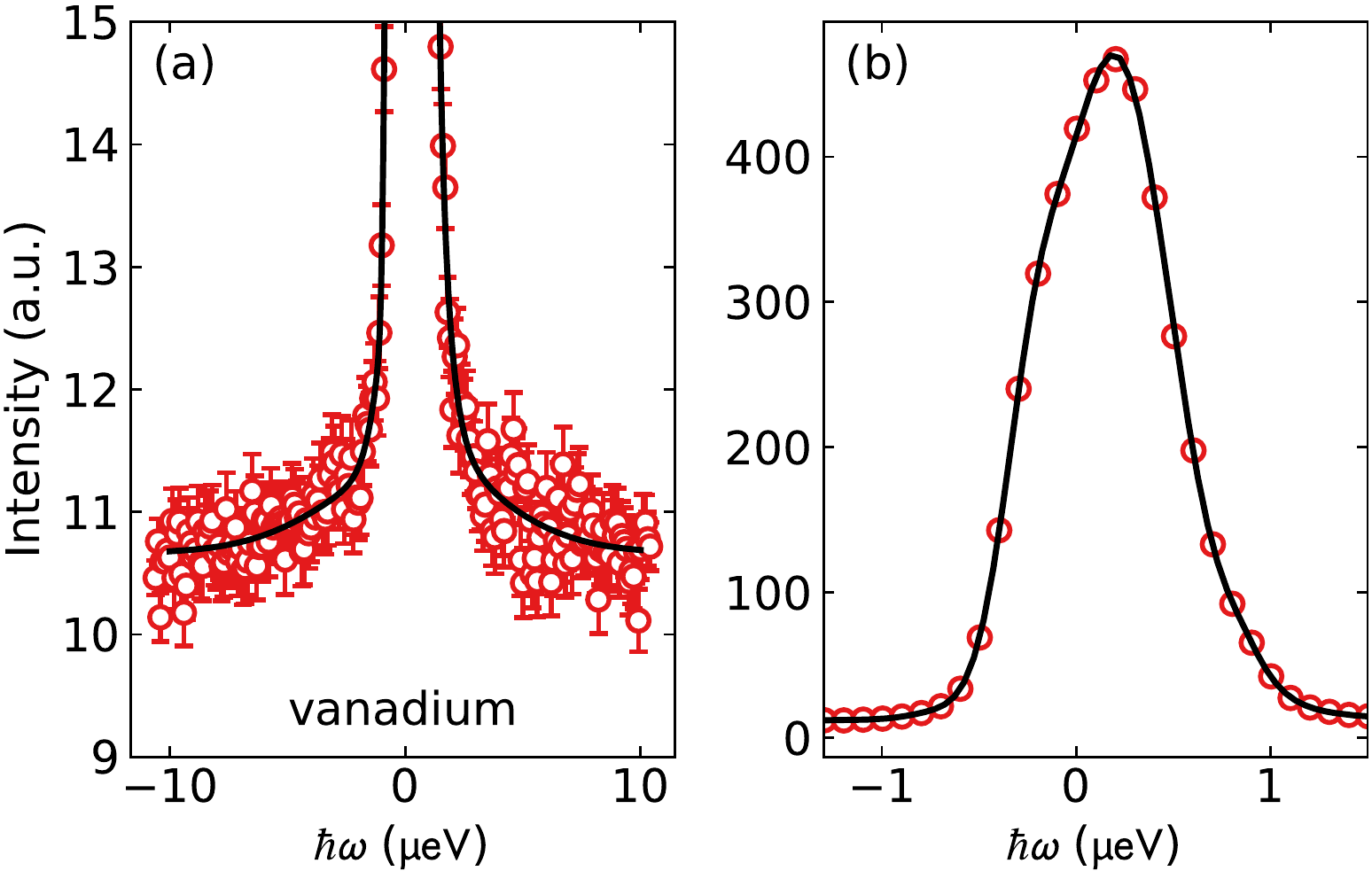}
	
	\caption{HFBS scattering from a vanadium standard sample, showing an asymmetric lineshape and our phenomenological fitted resolution function. (a) and (b) show different views of the same data.}
	\label{flo:vanadium}
	
\end{figure}

The resolution function for HFBS is asymmetric in energy \cite{HFBSpaper}, as is common for backscattering spectrometers \cite{Mamontov2011}. This is clearly shown by the incoherent elastic scattering acquired for a  vanadium standard sample in the $\pm 11 \> {\rm \mu eV}$ configuration shown in Fig. \ref{flo:vanadium}. There are a variety of instrumental causes for this asymmetry \cite{HFBSpaper}, and we account for it with a phenomenological resolution function that consists of five Gaussian peaks plus a Lorentzian peak:
\begin{equation}
R(\omega) = \sum_{i=1}^5 G(\omega, \omega_{0_i}, A_i, \sigma_i) + L(\omega, \omega_0, A_0, \Gamma_0).
\end{equation}
This function is normalized so $\int R(\omega) d \omega = 1$.
Here the peak centers $\omega_{0_i}$, areas $A_i$ and widths $\sigma_i$ are fitted sequentially: we started with a single Lorentizan fit, added a Gaussian term and refit, added another Gaussian and refit, etc. As shown in Fig. \ref{flo:vanadium}, this provides a very good reproduction of the lineshape. This normalized lineshape is used to fit all the hyperfine HFBS data in this paper, via convolution with models for the intrinsic physical spectrum.

\subsection{Hyperfine excitation model}\label{app:HyperfineModel}

The neutron cross section of powder-averaged nuclear hyperfine excitations is
\begin{equation}
\begin{aligned}
\Big( \frac{d^2 \sigma}{d \Omega d E} \Big)^{\pm} = & \frac{1}{3} \frac{k_f}{k_i} e^{-2W(Q)} \sum_i{ \rho_i I_i(I_i+1)} \\ 
 &   \frac{\sigma_i}{4 \pi} \Big[
	\sum_j p_{j}(T) \delta(\Delta_{M^{j}_{i}} - \Delta_{M^{j+1}_{i}} - \hbar \omega) \\
 & + 
\sum_j p_{j+1}(T) \delta(\Delta_{M^{j+1}_{i}} - \Delta_{M^{j}_{i}} - \hbar \omega) \Big].
\end{aligned}
\end{equation}
Here $\pm$ is positive and negative energy transfer, $\Delta_{M^j}$ are the hyperfine splitting energies, $\sigma_i$ is the Nd incoherent scattering cross section, $I$ is the nuclear spin state, $e^{-2W(Q)}$ is the Debye-Waller factor, and $k_i$ and $k_f$ are the incident and scattered neutron wave vectors. The sum $i$ is over isotopes with their relative abundance $\rho_i$ \cite{heidemann1970}, and the sum $j$ is over all energies. 
In this analysis, we approximate $\frac{k_f}{k_i} \approx 1$ because of the small energy transfers and $e^{-2W(Q)} \approx 1$ because of the low temperatures. The negative (positive) energy transfer scattering is suppressed (enhanced) by the Boltzmann weight $p(T) = \frac{e^{\mp \beta \Delta}}{e^{-\beta \Delta}+ e^{\beta \Delta}}$. 

The measured hyperfine peaks are broader than the resolution width defined above. We account for this with a Gaussian (which we use to model a random distribution of static moments) convolved with a Lorentzian (which we use to model finite lifetime of the excited nuclear hyperfine states), convoluted again with the resolution function. We apply the Boltzmann factor to the Gaussian distribution prior to convolution, because the nuclear states with large splitting will experience more thermal depopulation than those with smaller splitting (the effect of this is to suppress the negative energy tail of the $-\hbar \omega$ Gaussian and to enhance the positive energy tail of the $+\hbar \omega$ Gaussian).

The two stable isotopes of Nd with nuclear spin are 143 and 145, with relative abundances 12.2\% and 8.3\% respectively, and both with $I=7/2$.
Thus, the equation used for fitting in Fig. \ref{flo:HFBS_fit} is
\begin{equation}
\begin{aligned}
\Big( \frac{d^2 \sigma}{d \Omega d E} \Big)^{\pm} & =  \frac{1}{3} \bigg[ \frac{7}{2} \Big(\frac{7}{2}+1 \Big)\bigg] \\ &
\Big(0.122 \frac{\sigma_{143}}{4 \pi} \>
	\sum_j C(G,L,\Delta_{M^j_{143}} - \hbar \omega, T)  \\ & +
	0.083  \frac{\sigma_{154}}{4 \pi} \>
\sum_j C(G,L,\Delta_{M^j_{145}} - \hbar \omega, T)
	\Big)
\end{aligned}
\end{equation}
where $C(G,L,\Delta_M - \hbar \omega)$ is the convoluted broadening defined by a Gaussian width $G$ and a Lorentzian width $L$ with the Boltzmann weight defined by temperature $T$.

In fitting this model, we were also able to fit the effective temperature using the Boltzmann factor. Curiously, the lowest temperature data (which had a sample thermometer reading of $T=0.048$~K) gave a fitted temperature of $T=0.14(2)$~K. The difference is at least partially due to beam heating. Our previous measurements of $\rm Nd_3Sb_3Mg_2O_{14}$ (which, in total, had 1.3 times more absorption cross section in the beam) showed a Boltzmann fitted temperature of $T=0.10(3)$~K for a sample thermometer reading of 45 mK \cite{scheie2019homogenous}.

\section{Nuclear Schottky anomaly and the ordered moment}\label{app:NuclearSchottky}

The equation for a nuclear Schottky anomaly is 
\begin{equation}
C = \frac{1}{Z k_B T^2} \Big[
\sum_i E_i^2 e^{\frac{-E_i}{k_B T}} -
\frac{1}{Z} \big( \sum_i E_i e^{\frac{-E_i}{k_B T}} \big)^2
\Big]
\label{eq:NuclearSchottky}
\end{equation} \cite{scheie2019thesis}.
Here $E_i$ are the levels of the nuclear spin states, which are determined by the hyperfine nuclear spin Hamiltonian
\begin{equation}
{\cal H}  = a' I_z + P\big(I_z^2 - \frac{1}{3}I(I+1)\big)
\end{equation}
(see eq. 1 in ref. \cite{Bleaney1963}), where $a' = a \langle J_z \rangle$, $\langle J_z \rangle$ being the expectation value of the effective electronic spin $J$, and $P$ is a constant defining the hyperfine quadrupole coupling strength and the electronic quadrupole moment. For Nd$^{3+}$, the hyperfine quadrupole coupling strength is three orders of magnitude weaker than the dipole coupling \cite{Bleaney1963,aufmuth1992hyperfine}. Our CEF calculations indicate a potentially large static electric quadrupole order in the ground state doublet $\langle + | O^0_2 | + \rangle / [J(2J-1)] = 0.958$ (here $O^0_2$ is the quadrupole Stevens Operator \cite{Stevens1952} and $J$ is the total Nd$^{3+}$ spin). Even assuming a completely saturated quadrupole moment, the quadrupolar hyperfine level splitting only varies by $\approx1$\% relative to the dipole-only hyperfine splitting (0.014~$\rm \mu eV$ assuming the constants in ref. \cite{Bleaney1963} or 0.024~$\rm \mu eV$ assuming the constants in ref. \cite{aufmuth1992hyperfine}--two orders of magnitude smaller than the observed splitting). Thus, although we include it in our fits, the quadrupolar hyperfine coupling is negligible and the heat capacity is primarily from dipolar hyperfine effects. (These values also confirm that the inelastic neutron hyperfine peak broadening cannot be explained by Nd$^{3+}$ quadrupolar order.) In passing, we note that it is also possible for nuclear levels to be split by crystal electric fields, but the nuclear quadrupole moment is small for Nd, and these effects tend to be <0.1~$\mu$eV for Nd \cite{Chattopadhyay1999}; therefore we do not consider these effects in our analysis.

The high temperature limit of Eq. \ref{eq:NuclearSchottky} is $C \propto \frac{1}{T^2}$. 
If the ordered magnetic moment size varies from site to site but is the same ion, the high-temperature tail of the nuclear Schottky anomaly gives the root-mean-squared ordered moment. This can be demonstrated with a sum over all sites indexed by $k$ in the high-temperature limit:
\begin{equation}
C_{net} \propto \sum_k \frac{E_k^2}{T^2} \propto  \frac{ \sum_k \langle J_{z_k}^2 \rangle}{T^2} = \frac{ \langle J_{z_{net}}^2 \rangle}{T^2} 
\end{equation} 
and thus
\begin{equation}
J_{z_{net}} = \sqrt{\sum_k \langle J_{z_k}^2 \rangle }.
\label{eq:rms_moment}
\end{equation}
However, this approximation is only valid in the high-temperature limit. As one nears the Schottky peak, this approximation breaks down. If the distribution is very broad, the peak height itself will be suppressed. In practice, the ordered moment fitted from the upturn will be in between the mean and RMS moment.

We finally note that there is an intrinsic disorder associated with the different nuclear spins of the 7 different isotopes of Nd, which could play a role in the low temperature electronic magnetism. Experiments on nuclear spin free isotopic samples would be of interest to look into this. 

\section{Heat capacity measurements}

\begin{figure}[H]
	\centering\includegraphics[width=0.44\textwidth]{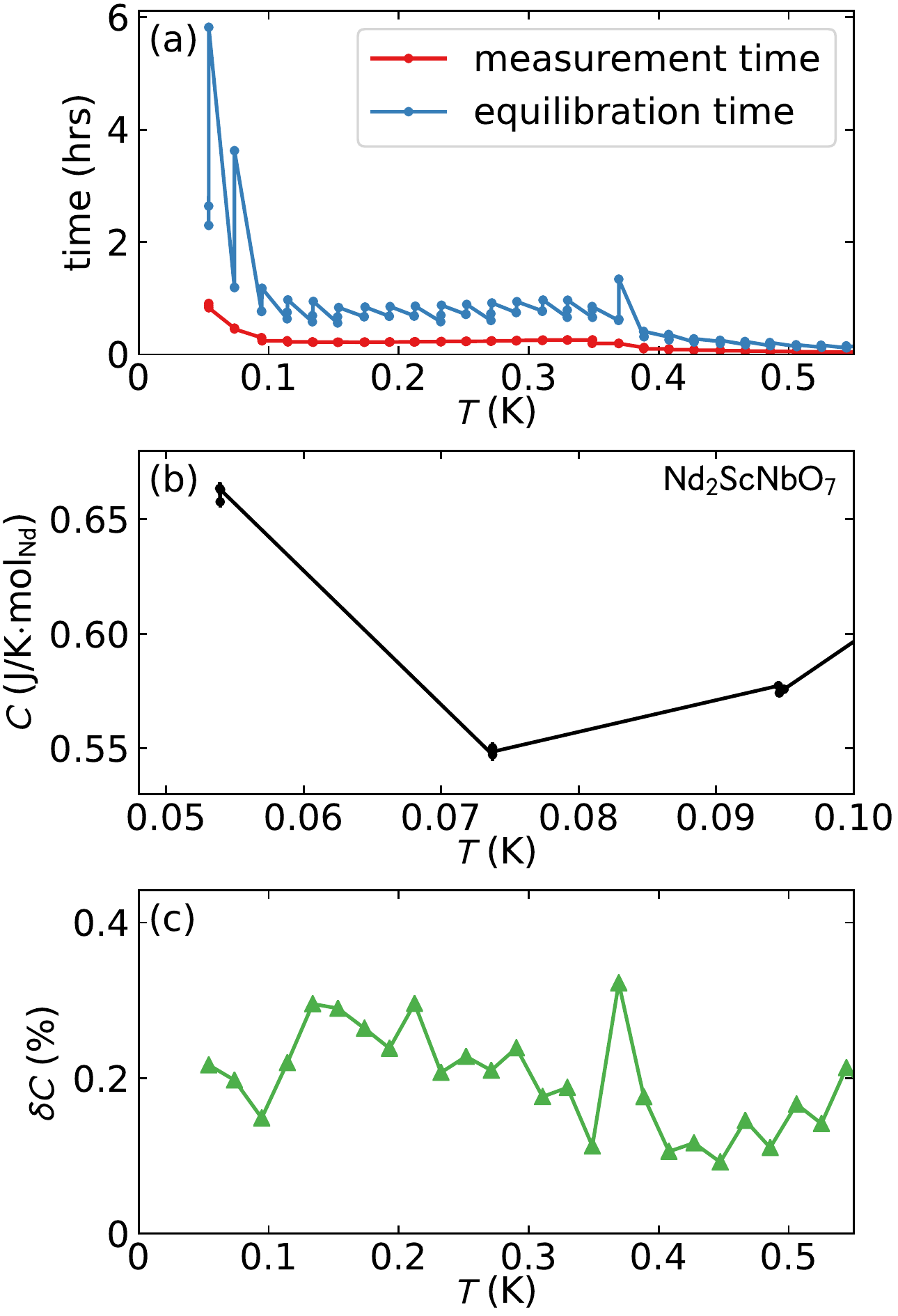}
	
	\caption{(a) Measurement and equilibration times for the heat capacity measurements of $\rm{Nd_2ScNbO_7}$. The lowest temperature data points had a measurement time of $\sim 1~$hr and equilibration times of several hours. (b) Three lowest heat capacity temperatures measured, showing the small spread in measured heat capacity and temperature for each data point. (c) Uncertainty $\sigma_{net}$ for heat capacity, as computed with eq. \ref{eq:NetUncertainty}.}
	\label{flo:HC_details}
	
\end{figure}

Heat capacity data were acquired using  a Quantum design PPMS with the semi-adiabatic relaxation method, wherein the time dependent thermal response $T(t)$ of the sample stage to a heat pulse applied to it is monitored. With knowledge of the thermal conduction between the sample stage and the thermal reservoir the specific heat of the sample is determined from a fit to $T(t)$ \cite{PPMS_Manual}. 
We applied sufficient  heat to achieve a 3\% temperature rise at each temperature (ie, the 95 mK data point had a $\sim 3$ mK temperature rise). We let the temperature equilibrate between each data point until the sample temperature was stable to within 3\% of the temperature rise over the measurement time for three consecutive measurements (equilibration at each point was at least three times the measurement time---see Fig. \ref{flo:HC_details}).

\subsection{Temperature calibration}

Whenever one observes an unusual experimental signal (as we have in this paper), it is always important to consider errors in the measurement technique.
If temperature calibration were off by 100\% at 100 mK (showing 100 mK when the sample is really at 200 mK), this positive $\gamma$ term would disappear. However, we consider this to be unlikely for four reasons:
(a) The transition temperature $T_N = 370(10)~{\rm mK}$ measured from heat capacity is in agreement with $T_N = 371(2)~{\rm mK}$ from neutron diffraction, suggesting that the low-temperature calibration is correct to within 10~mK at 300 mK. (b) The compound measured immediately prior to this on the same PPMS was $\rm Nd_3Sb_3Mg_2O_{14}$, which gave a Schottky anomaly upturn beautifully consistent with the ordered moment from neutron measurements \cite{scheie2019homogenous}, indicating good calibration. The same equipment and calibration was used here. (c) Sample coupling---the measure of the exponential nature of the semi-adiabatic heating and cooling curves on a PPMS---is $>95$\% for the entire temperature range, indicating excellent exponential-like heating and cooling behavior and reliable heat capacity values.
(d) The same low-temperature density of states was observed in $\rm Nd_2Zr_2O_7$ heat capacity (which was measured with a completely different experimental setup), suggesting an intrinsic behavior to  Nd$^{3+}$ pyrochlores.
Therefore, we consider the positive offset in $C/T$ to be a reliable experimental results subject to the statistical error bars quoted.

\subsection{Uncertainty}

We measured heat capacity at each temperature three times, and this repetition revealed good temperature stability and very reproducible values for every temperature measured [Fig. \ref{flo:HC_details}(b)].
We combined the three data points by taking the mean of heat capacity and temperature. Uncertainty in heat capacity for each data point was computed as the standard error of the mean of the three data  points $\sigma_{std}/\sqrt{n}$ added in quadrature to the standard error of the weighted mean $(\sum_i \sigma_i^{-2})^{-1/2}$ (this is to account for both the uncertainties generated by the PPMS fitting routine values plus the spread of data points) so that uncertainty is
\begin{equation}
\sigma = \sqrt{\frac{\sigma_{std}^2}{3} + \Big(\sum^3_i \sigma_i^{-2} \Big)^{-\frac{1}{2}}}.
\end{equation}
Here $\sigma_{std}$ is the standard deviation of the three data points, and $\sigma_i$ are the uncertanties of the individual data points (which comes from covariance matrix of the exponential fits). 
Uncertainty in temperature was computed as the standard error of the mean of the three averaged points. For every temperature the uncertanties are quite small: $\sigma_T<0.25$~mK and $\sigma_c<0.3\%$ below 1 K. Error bars are always smaller than the data point markers in the main text, and these uncertanties are used to define the $\chi^2$ of the different fitted models.

The fitting algorithm used for heat capacity models (Scipy's "curve\_fit" \cite{scipy}, which uses a least-squares method) does not account for uncertainty in the $x$-axis (in this case temperature), so we converted $\sigma_T$ to $\sigma_C$ by multiplying the uncertainty in temperature $\sigma_T$ by the slope of the data at that point $s_i$. We then added it in quadrature to the heat capacity uncertainty:
\begin{equation}
{\sigma_{net}}_i = \sqrt{{\sigma_C}_i^2 + ({\sigma_T}_i s_i)^2}.
\label{eq:NetUncertainty}
\end{equation}
These $\sigma_{net}$, plotted in Fig. \ref{flo:HC_details}(c),  were used in defining $\chi^2$ for the fits.

\subsection{Silver subtraction}

\begin{figure}
	\centering\includegraphics[width=0.44\textwidth]{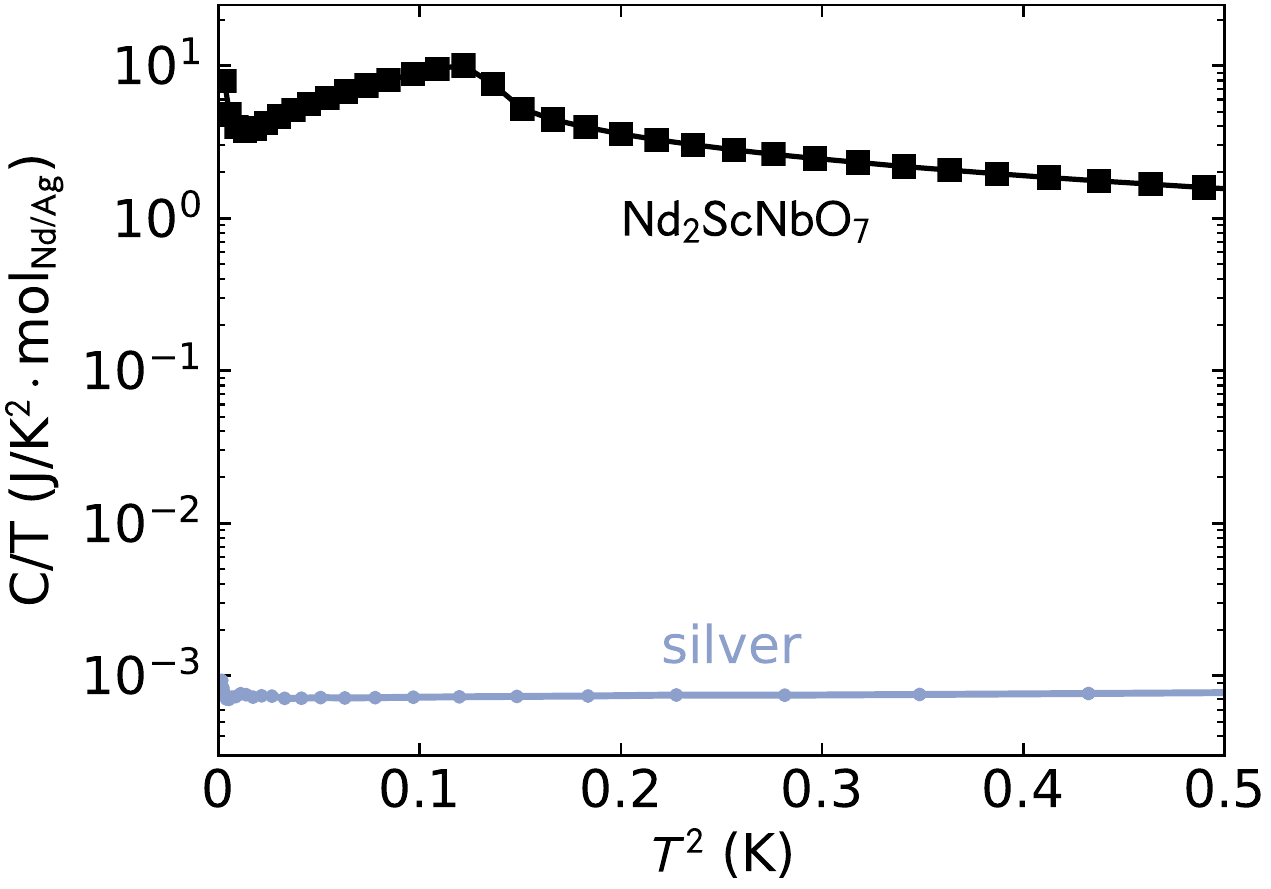}
	
	\caption{Heat capacity of $\rm{Nd_2ScNbO_7}$ compared to the silver powder it was mixed with. The difference in heat capacity is well over three orders of magnitude. }
	\label{flo:silverHC}
	
\end{figure}

For the heat capacity measurement, $\rm{Nd_2ScNbO_7}$ was mixed with silver powder with a 1:1 mass ratio, and then the contribution from silver was subtracted from the data post-facto. Because the heat capacity of silver is so small below 1 K (three orders of magnitude below $\rm{Nd_2ScNbO_7}$) the effect of this subtraction is very slight as shown in Fig. \ref{flo:silverHC}.
This figure also shows the Sommerfeld coefficient of silver to be $\gamma = 7.1(2)\times 10^{-4} {\rm \frac{J}{mol \> K^2}}$, which is far too small to account for the large $C/T$ offset in low temperature $\rm{Nd_2ScNbO_7}$ heat capacity.

\section{Heat capacity fits}\label{app:HCmodelFits}

\subsection{Models}

As noted in the main text, we fit the heat capacity using three different models. In this section we list the best fit parameters for each model.

\subsubsection{Dispersive magnons + flat band}

This model is given in Eq. eq. \ref{eq:MagnonModel} in the main text
$$
C = A \> C_{flat} + C_{ld}(\Delta, c) + C_{schottky}(\mu),
$$
which includes a a flat band excitation (eq. \ref{eq:EinsteinMode}, with $\hbar \omega_0 = 0.07$~meV) plus a gapped linear dispersive spin wave mode plus a nuclear Schottky anomaly. The calculation for the linear dispersive mode heat capacity $C_{ld}(\Delta, c)$ with spin wave velocity $c$ is based off ref. \cite{scheie2019homogenous}.
Because we do not know the precise spin wave spectrum of $\rm Nd_2ScNbO_7$, we allow the relative contributions of the flat band and the spin wave mode to be fitted parameters. (This is equivalent to fitting the phonon specific heat to the sum of an Einstein mode and a Debye spectrum.)
The best fit parameters and reduced $\chi^2$ are given in Table \ref{tab:Magnon+FlatBand_BestFit}.

\begin{table}[H]
	\caption{Best fit parameters for eq. \ref{eq:MagnonModel} }
	\begin{ruledtabular}
		\begin{tabular}{c|c c c c}
			$\chi^2$ & $A$ 	& $\Delta$ (meV) 	& $c$ (m/s) 	& $\mu$ ($\mu_B$) \\
			\hline
			3896 	  & 0.40(12)	& 0.00(0)	& 24.2(1.3)	& 1.72(8) 
	\end{tabular}\end{ruledtabular}
	\label{tab:Magnon+FlatBand_BestFit}
\end{table}

\subsubsection{Power law + flat band}

This model is given in eq. \ref{eq:PowerLawModel} in the main text
$$
C = A \> C_{flat} + B \> T^n + C_{schottky}(\mu),
$$
which includes a flat band excitation plus a spin wave mode with an arbitrary power plus a nuclear Schottky anomaly. 
The best fit parameters are given in Table \ref{tab:PowerLaw+FlatBand_BestFit}. The weight on the flat band excitation converged to zero. This indicates that this fit is not very reliable, because neutron scattering clearly shows the flat band to be quite intense (at least $17$\% of the measured spectral weight) and not a tiny contribution like the fit indicates.

\begin{table}[H]
	\caption{Best fit parameters for eq. \ref{eq:PowerLawModel}}
	\begin{ruledtabular}
		\begin{tabular}{c|c c c c}
			$\chi^2$ & $A$ 	& $B$ 	& $n$ 	& $\mu$ ($\mu_B$) \\
			\hline
			140 	  & 0.00(4)	& 49(3)	& 2.13(3)	& 1.32(3) 
	\end{tabular}\end{ruledtabular}
	\label{tab:PowerLaw+FlatBand_BestFit}
\end{table}

\subsubsection{Dispersive magnons + linear term}

This model is given in Eq. \ref{eq:LinearTermModel} in the main text
$$
C = A \> C_{flat} + C_{ld}(\Delta=0, c) + \gamma T + C_{schottky}(\mu)
$$
which includes a flat band plus gapless spin wave mode plus a nuclear Schottky anomaly. 
The best fit parameters are given in Table \ref{tab:Magon+LinearTerm_BestFit}.

\begin{table}[H]
	\caption{Best fit parameters for eq. \ref{eq:LinearTermModel}}
	\begin{ruledtabular}
		\begin{tabular}{c|c c c c}
			$\chi^2$ & $A$ 	& $c$ (m/s)	& $\gamma$ 	& $\mu$ ($\mu_B$) \\
			\hline
			7.58 	  & 0.072(7)	& 26.62(11)	& 3.23(4)	& 1.130(8) 
	\end{tabular}\end{ruledtabular}
	\label{tab:Magon+LinearTerm_BestFit}
\end{table}

We also tried this fit allowing the gap $\Delta$ to be nonzero, but the value converged to zero every time. Thus we took it out of the fit and set $\Delta=0$ so that the number of parameters matched Eq. 2.

\subsubsection{Dispersive magnons + $T^{3/2}$ term}

\begin{figure}
	\centering\includegraphics[width=0.44\textwidth]{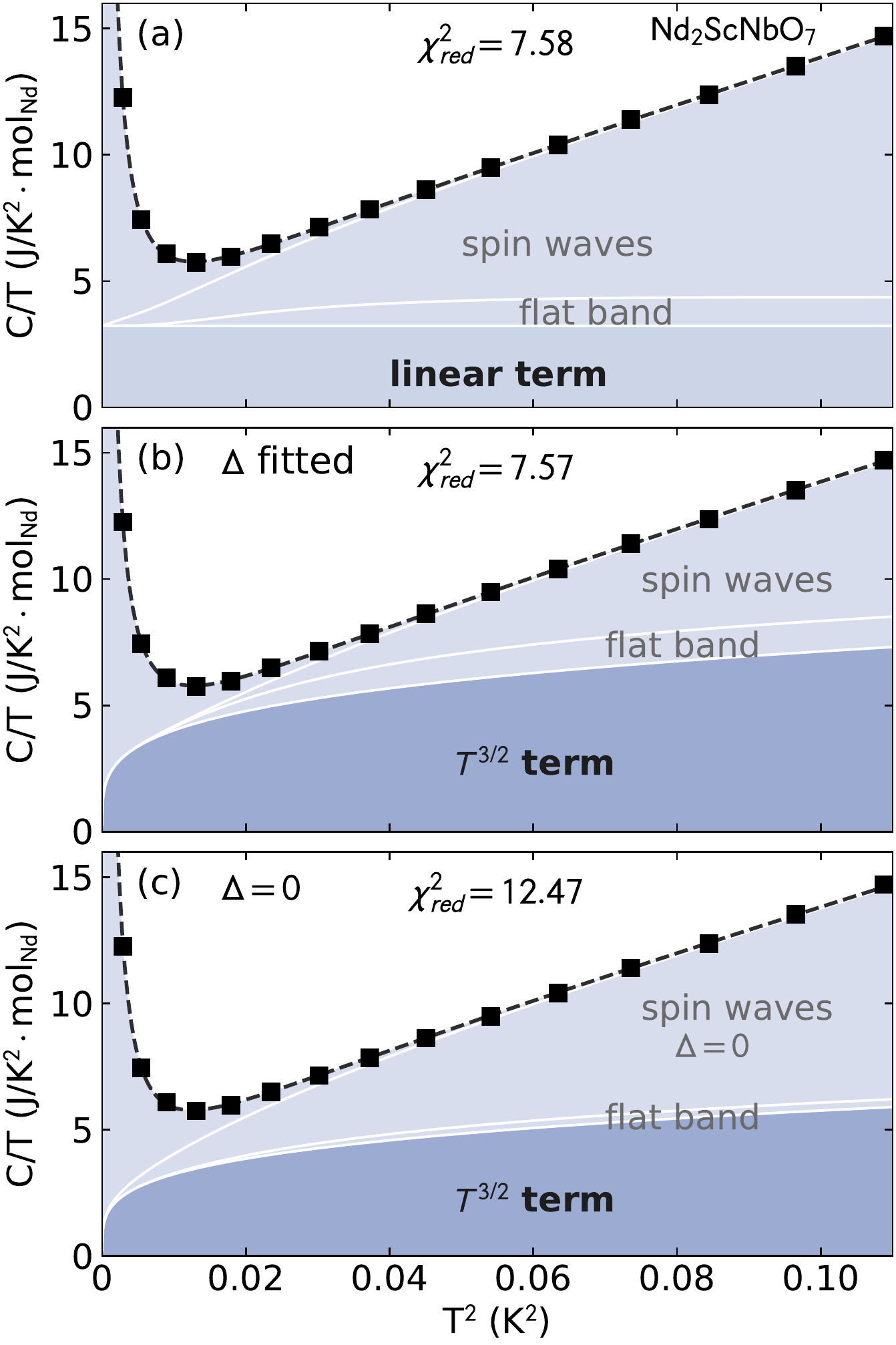}
	
	\caption{Low temperature heat capacity of $\rm Nd_2ScNbO_7$ fitted to (a) a model including a linear term, (b) a model including a $T^{3/2}$ contribution from spinons and a gapped magnon mode, and (c)  a model including a $T^{3/2}$ contribution and a gapless ($\Delta=0$) magnon mode. All models match the data well, but as revealed by $\chi^2$ comparisons, the gapped magon with $T^{3/2}$ heat capacity matches the best.}
	\label{flo:Alt_HC_fits}
\end{figure}

It was theoretically predicted in ref. \cite{Liu_2019} that a magnetically ordered pyrochlore can have spinons excitations which produce a $T^{3/2}$ power law dependence in heat capacity.
To test this, we fitted the model in eq. \ref{eq:T3/2model} two ways: once allowing $\Delta$ to be a free fitted parameter, and once fixing $\Delta=0$ (to keep the same number of fitted parameters as eq. \ref{eq:MagnonModel}).
The results are plotted in Fig. \ref{flo:Alt_HC_fits} and the best fit parameters are in Table \ref{tab:Magon+T3/2_BestFit}.
This model fits as well as the linear term model, suggesting that the anomalous low-energy density of states could be due to the $C \propto T^{3/2}$ spinon excitations considered in ref. \cite{Liu_2019}.
This model also predicts a magnon gap of 0.07 meV, which is almost exactly the energy of the flat band---suggesting that this model could be a better reflection of the actual magnon behavior in $\rm Nd_2ScNbO_7$.

\begin{table}[H]
	\caption{Best fit parameters for eq. \ref{eq:T3/2model}. The first top row gives the best fits parameters for when the magnon gap was a fitted parameter, the bottom row gives the best fit parameters for gapless magnons.}
	\begin{ruledtabular}
		\begin{tabular}{c|c c c c c}
			$\chi^2$ & $A$ 	& $\Delta$	& $c$ (m/s) 	& $B$ 	& $\mu$ ($\mu_B$) \\
			\hline
			7.57 	  & 0.077(4)	& 0.070(16)	& 27.9(8)	& 12.7(18)	& 1.178(8) \\
			12.47 	  & 0.02(1)		& 0.0		& 28.5(18)	& 10.2(16)	& 1.216(9) 
	\end{tabular}\end{ruledtabular}
	\label{tab:Magon+T3/2_BestFit}
\end{table}

\subsection{Varying maximum fitted temperature}

\begin{figure*}
	\centering\includegraphics[width=0.75\textwidth]{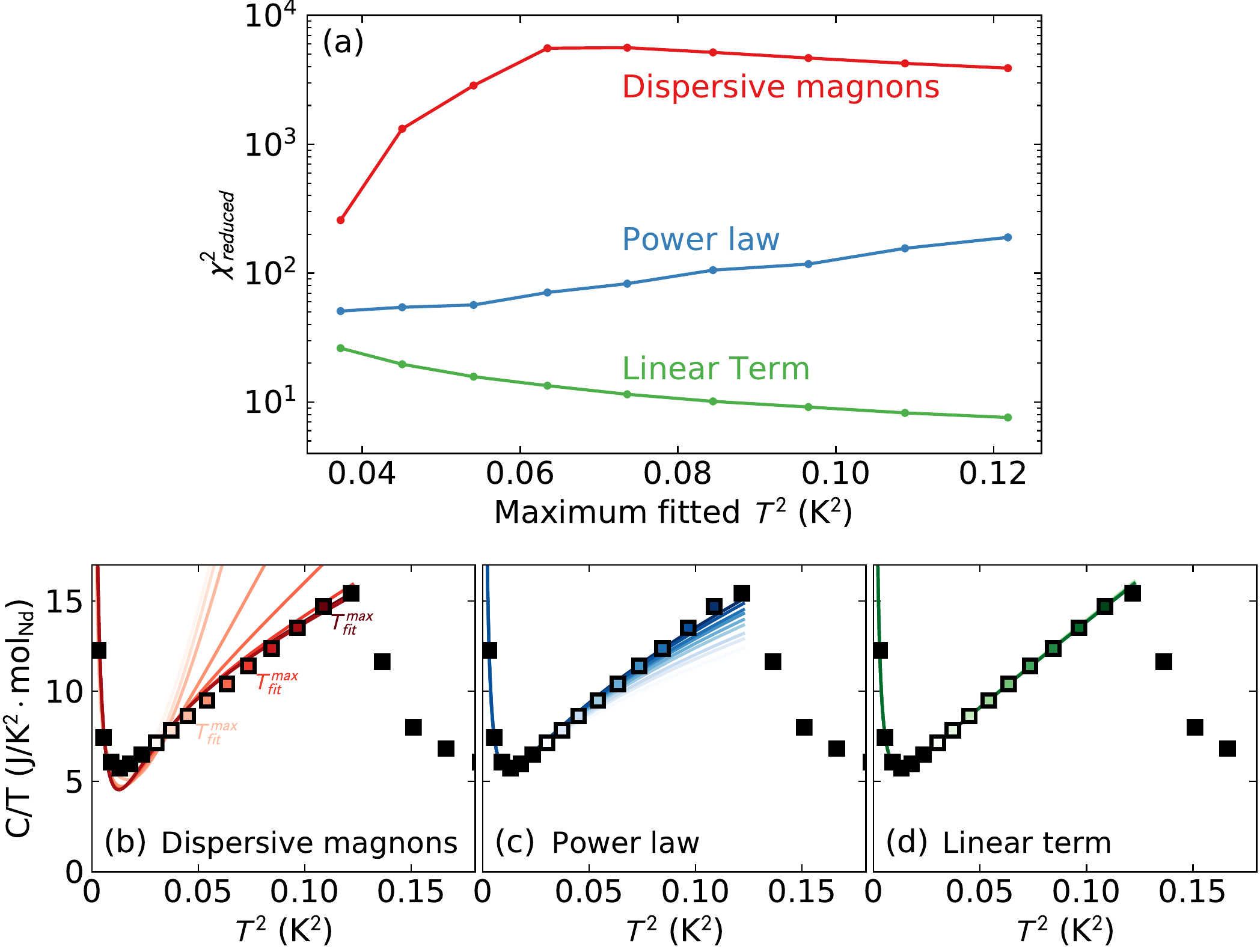}
	
	\caption{(a) reduced $\chi^2$ for the three models as a function of maximum fitted temperature. (b) Dispersive magnon model fits (solid lines) compared with data. The shade of the line indicates the maximum fitted temperature, which is indicated by the shade of the data marker. (c) Power law model fits (solid lines) compared with data. (c) Linear term model fits (solid lines) compared with data.}
	\label{flo:HC_fit_varyTmax}
\end{figure*}

The range of data fitted for each different model was $T < 0.33$~K. This temperature was chosen because it is the highest temperature at which the heat capacity exhibits roughly $T^3$ behavior. If we decrease the highest fitted temperature, the fits give the same general results: the model including only dispersive magnons plus a flat band fits the data poorly, a power law fits the data better, and a model including a linear heat capacity term fits the data best. 

Figure \ref{flo:HC_fit_varyTmax} illustrates this by plotting reduced $\chi^2$ as a function of maximum fitted $T$. Any reasonable maximum fitted $T$ still causes the linear term model to fit the data best by over an order of magnitude. Curiously, the fitted temperature range makes almost no difference for the fitted values for the linear term model, as demonstrated in Fig \ref{flo:HC_fit_varyTmax}(d). 
Thus, the main conclusions of this paper are independent on which temperature range is chosen for a fit.

\section{Monte Carlo charge ice simulations}\label{app:MonteCarlo}

We performed Monte Carlo (MC) simulations with a large box of $6\times6\times6$ pyrochlore unit cells, allowing the Nb and Sc sites to flip from one to another.
Charge-ice correlations were simulated by assigning an energy $-J$ to every Nb-Sc pair on a tetrahedron and $+J$ to every Nb-Nb or Sc-Sc pair on a tetrahedron. This led to a two-Nd-two-Sc tetrahedron having an energy $-2J$, a three-Nd-one-Sc or three-Sc-one-Nd tetrahedron having an energy of $0J$, and a four-Nd or four-Sc tetrahedron having an energy $+6J$. The lowest energy state is a "charge ice", with two Nd and two Sc on each tetrahedron, but simulations at finite MC temperature will deviate from this.
 We used a Metropolis algorithm to calculate the relative frequency of the 13 symmetry-unique local environments as a function of temperature, sweeping through the lattice four times between measuring and taking 250 samples at each temperature. The percentage of sites with $D_3$ symmetry as a function of MC temperature is shown in Fig. \ref{flo:ChargeIceMC}, and the code for these MC simulations can be found in the Supplemental Information \cite{SuppMat}.

\begin{figure}[H]
	\centering\includegraphics[width=0.49\textwidth]{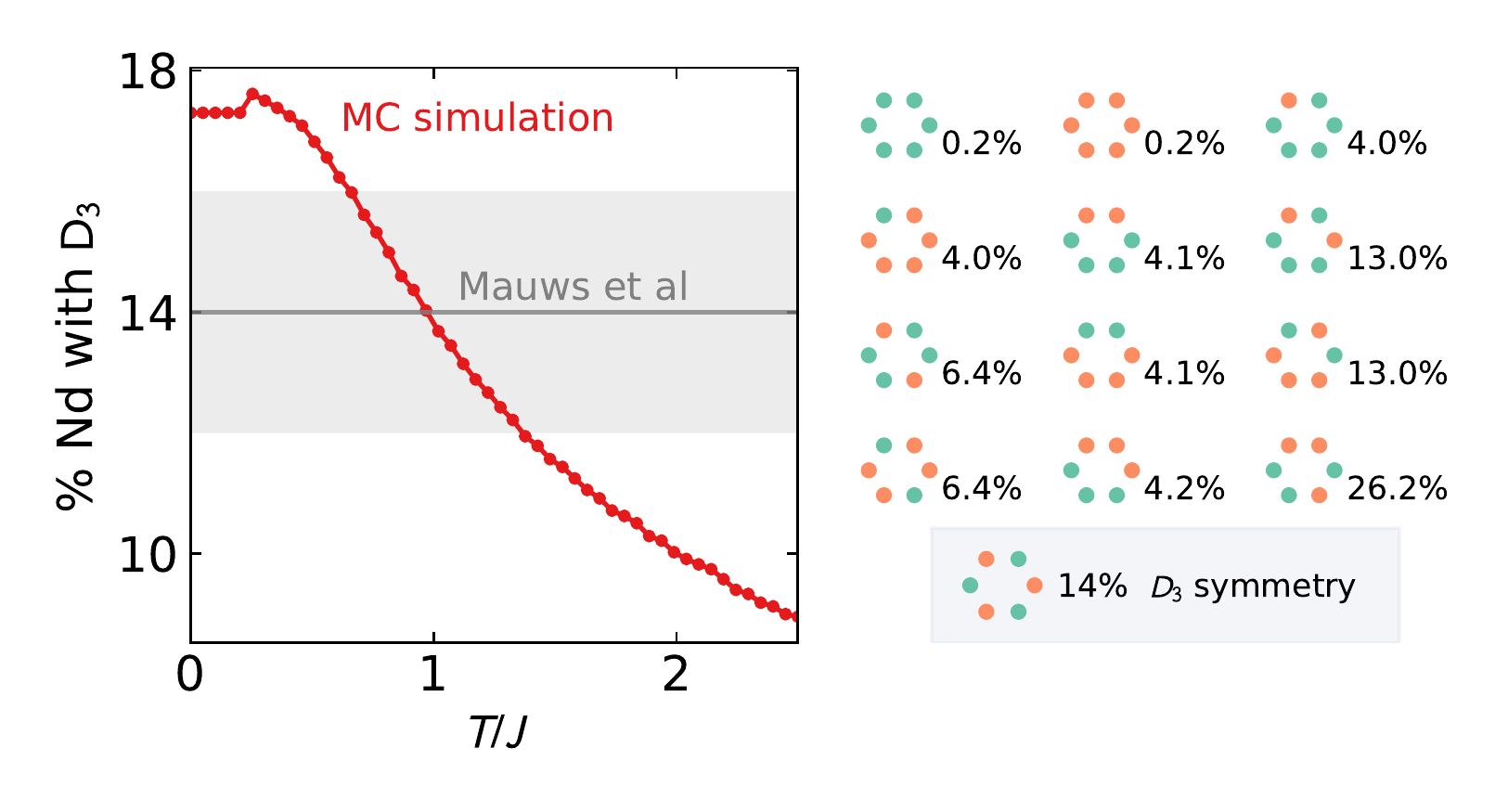}
	
	\caption{The left panel shows the percentage of Nd sites with $D_3$ symmetry as a function of MC temperature. 14\% of sites have $D_3$ symmetry at approximately $T/J=1$. (The low-temperature jump is a finite-size effect and varies as the system size is changed.) The right panel shows the relative populations of the 13 symmetry-unique environments for Nd in $\rm Nd_2ScNbO_7$ where the orange dots are Sc and the blue-green dots are Nb.}
	\label{flo:ChargeIceMC}
	
\end{figure}

We then calculated a point charge model using PyCrystalField \cite{PyCrystalField} for every symmetry-unique ring around the Nd sites (there are $2^6$ configurations but only thirteen symmetry-unique rings). The point-charge model is an approximation, but it is usually close enough to give a qualitative picture of the CEF Hamiltonian \cite{Hutchings1964,EDVARDSSON1998}.
To account for the different charges of the Nd and Sc sites, we used different effective charges for each and assumed that the O$^{2-}$ ligands shift towards the Nb$^{5+}$ by 3\% as compared to Sc$^{3+}$.
To constrain the model to fit better with the observed CEF transitions in ref. \cite{mauws2019order}, we fit the effective charges in the point charge model to minimize the difference between eigenvalues of the $D_3$ symmetric sites and the observed transitions in ref. \cite{mauws2019order}. This fit yielded effective charges of $-0.77e$ and $-2.32e$ on the two symmetry-independent O sites,  $2.71e$ on the Sc site, and  $6.39e$ on the Nb site. We then used these effective charges to simulate the CEF Hamiltonian of all 13 symmetry-unique ligand environments for Nd$^{3+}$, as shown in Fig. \ref{flo:CEF_simulation}.

\begin{figure}
	\centering\includegraphics[width=0.47\textwidth]{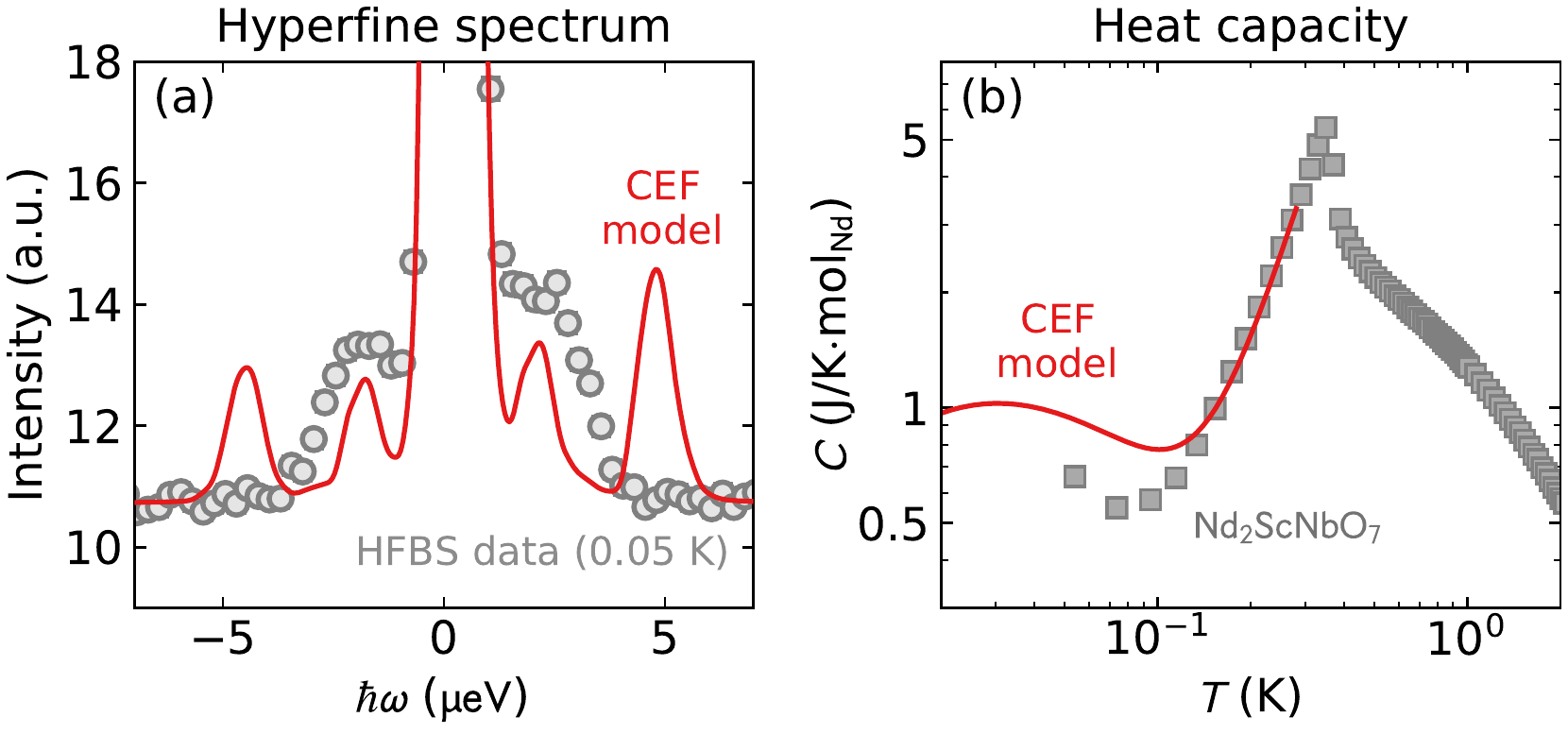}
	
	\caption{Calculated hyperfine spectrum (a) and heat capacity (b) from the CEF disorder point-charge model. The CEF spectrum predicts some sites to have much larger static moments than is observed, and the distribution is not nearly as continuous as in experiment. Heat capacity also shows the CEF model overestimates the mean ordered moment.}
	\label{flo:CEF_calculatedSpectrum}
	
\end{figure}

To test the correspondence to the experimental data, we calculated the hyperfine spectrum and nuclear Schottky heat capacity, as shown in Fig. \ref{flo:CEF_calculatedSpectrum}. This shows a mostly bimodal distribution of static moments in the hyperfine spectrum, which is not true of the actual data. We also (unsurprisingly) see that the nuclear Schottky anomaly onset is at a much higher temperature than experiment. These discrepancies are partially due to the simplistic assumptions we have made in modeling the moments with a point-charge model, and partially due to inter-site exchange which affect the static moments size---either singlet formation or collective fluctuations.

\section{Calculated $\rm Nd_2Zr_2O_7$ heat capacity}\label{app:HeatCapacityFromSpinWaves}

We calculated heat capacity by simulating the $\rm Nd_2Zr_2O_7$ spin wave spectrum from the Hamiltonian in ref. \cite{Xu_2019_Anisotropic} using the SpinW software package \cite{SpinW}. We simulated the spin wave spectrum over the reciprocal space range $0<h<1 \> RLU$, $0<k<1 \> RLU$, and $0<l<1 \> RLU$, and then calculated heat capacity following ref. \cite{Maestro_2004_HC} as
\begin{equation}
C_v = \beta^2 \sum_{\bf k} \sum_{\alpha} [\eta_{\alpha}({\bf k}) n_B(\eta_{\alpha}({\bf k})) ]^2 \exp[\beta \eta_{\alpha}({\bf k})]
\end{equation}
where $n_B(\eta_{\alpha}({\bf k}))$ is the Bose factor. We then divided the calculated heat capacity by the number of spin wave modes to get heat capacity per Yb ion, and then divided again by six because there are six symmetry-related regions in the box chosen. (This is a computationally expedient alternative to summing over only the first Broullin zone, but is formally equivalent.)
The results are plotted as the red line in Fig. \ref{flo:HC_comparison}, and capture the higher temperature non-$T^3$ behavior well.

Figure \ref{flo:HC_comparison} shows data from four compounds: $\rm Nd_2ScNbO_7$, $\rm Nd_2Zr_2O_7$ \cite{Blote1969}, $\rm Nd_3Sb_3Mg_2O_{14}$ \cite{MyPaper} (not a pyrochlore but a derivative of the pyrochlore lattice sharing many of its features), and $\rm Nd_2Sn_2O_7$ \cite{Blote1969}. The light-blue line shows the raw data, and the dark blue line shows the data with a phenomenological Schottky anomaly subtracted based on a fitted power-law $T^n$ for electronic heat capacity.
The latter two ($\rm Nd_3Sb_3Mg_2O_{14}$ and $\rm Nd_2Sn_2O_7$) are conventional ordered magnets, showing $T^3$ heat capacity from linear dispersive spin waves. Slight negative offsets ($y$-intercept < 0) indicate small spin wave gaps of $35(1)~\rm \mu eV$ and $101(7)~\rm \mu eV$, respectively.

%


\begin{thebibliography}{84}
\bibitem [{NIS()}]{NIST_disclaimer}%
	\BibitemOpen
	\href@noop {} {}\bibinfo {note} {Certain commercial instruments are
		identified in this paper to foster understanding. Such identification does
		not imply recommendation or endorsement by the National Institute of
		Standards and Technology, nor does it imply that the instruments identified
		are necessarily the best available for the purpose.}\BibitemShut {Stop}%
\end{thebibliography}

\end{document}